\documentclass[symmetry,article,accept,moreauthors,pdftex]{mdpi} 

\usepackage{latexsym}
\usepackage{amsmath}
\usepackage{amssymb}
\usepackage{amsfonts}

\usepackage{placeins}
\usepackage{supertabular}

\usepackage[mathscr,scaled=1.15]{urwchancal}
\DeclareFontFamily{OT1}{pzc}{}
\DeclareFontShape{OT1}{pzc}{m}{it}%
{<-> s * [1.15] pzcmi7t}{}
\DeclareMathAlphabet{\mathpzc}{OT1}{pzc}{m}{it}

\firstpage{1} 
\pubvolume{xx}
\issuenum{1}
\articlenumber{5}
\pubyear{2021}
\copyrightyear{2021}
\history{Received: date; Accepted: date; Published: date}

\usepackage{bm}
\usepackage{slashed}
\def\beq{\begin{equation}}
\def\eeq{\end{equation}}
\def\bea{\begin{eqnarray}}
\def\eea{\end{eqnarray}}
\def\nn{\nonumber}

\def\bx{{\bm x}}
\def\by{{\bm y}}
\def\bz{{\bm z}}

\def\bq{{\bm q}}
\def\bp{{\bm p}}

\def\bpi{\bm{\pi}}
\def\btau{{\bm \tau}}

\def\sla{\slashed}

\Title{Nonequilibrium Dynamics of the Chiral Quark Condensate under a Strong Magnetic Field}

\newcommand{\orcidauthorA}{0000-0003-1713-8578} 
\newcommand{\orcidauthorB}{0000-0003-4185-8356} 

\Author{Gast\~ao Krein$^{1,\ddagger}$\orcidA{}
and 
Carlisson Miller$^{1,\ddagger}$\orcidB{}
}

\AuthorNames{Gast\~ao Krein\orcidauthorA{} and Carlisson Miller\orcidauthorB{} }

\address{%
$^{1}$ \quad 
Instituto de F\'{\i}sica Te\'orica, Universidade Estadual Paulista,
Rua Dr. Bento Teobaldo Ferraz, 271 - Bloco II, 01140-070 S\~ao Paulo, SP, Brazil}

\corres{Correspondence: gastao.krein@unesp.br}

\secondnote{These authors contributed equally to this work.}

\abstract{
Strong magnetic fields impact quantum-chromodynamics (QCD) properties in several situations; 
examples include the early universe, magnetars, and heavy-ion collisions. These examples share a 
common trait: time evolution. A prominent QCD property impacted by a strong magnetic field is the 
quark condensate, an approximate order parameter of the QCD transition between a high-temperature 
quark-gluon phase and a low-temperature hadronic phase. We use the linear sigma model with quarks to 
address the quark condensate time evolution under a strong magnetic field. We use the closed time 
path formalism of nonequilibrium quantum field theory to integrate out the quarks and obtain a mean-
field Langevin equation for the condensate. The Langevin equation features dissipation and noise 
kernels controlled by a damping coefficient. We compute the damping coefficient for magnetic field 
and temperature values achieved in peripheral relativistic heavy-ion collisions and solve the 
Langevin equation for a temperature quench scenario. The magnetic field changes the dissipation and 
noise pattern by increasing the damping coefficient compared to the zero-field case. An increased 
damping coefficient increases fluctuations and time scales controlling condensate's short-time 
evolution, a feature that can impact hadron formation at the QCD transition. The formalism 
developed here can be extended to include other order parameters, hydrodynamic modes, and 
system's expansion to address magnetic field effects in complex settings as heavy-ion collisions, 
the early universe, and magnetars.}

\begin{document}

%
%

\section{Introduction}
\label{sec:intro}

Strong magnetic fields impact prominent quantum-chromodynamics (QCD) phenomena, 
notably those associated with QCD's approximate chiral symmetry in the light-quark 
sector. Special in this respect is the impact on the chiral condensate, as revealed 
by recent lattice QCD calculations~\cite{DElia:2012ems,Endrodi:2014vza,Ding:2020pao}. 
The~chiral~condensate is an approximate order parameter for the finite 
temperature QCD transition between a high-temperature quark-gluon phase (QGP) and 
a low-temperature hadronic phase. The~transition likely qualifies as a crossover 
(not a phase transition), in that the chiral condensate is nearly zero in~the~QGP 
phase, and nonzero in the hadronic phase, with a rapid change~(not~a~jump) around 
the~pseudocritical temperature $T_{\rm pc} \simeq 150$~MeV~\cite{Aoki:2006we}. Such 
a rapid change in the condensate's value is key to our understanding of how 
protons and neutrons (and other light-flavor hadrons) acquire their masses from 
almost massless quarks and gluons~\cite{Wilczek:2008zz,Roberts:2020hiw}. 
Phenomenologically, QCD matter under strong magnetic fields occurs in different 
settings, to name three of great current interest: the early 
universe~\cite{Vachaspati:1991nm,Grasso:2000wj}, magnetars~\cite{Kouveliotou:1998ze,
Duncan:1992hi}, and relativistic heavy-ion collisions~\cite{Rafelski:1975rf,
Kharzeev:2007jp}. Magnetized~QCD~matter in~those~settings evolves in time, albeit 
under very different time scales. A~QGP~to~hadron transition occurring under 
such circumstances typifies a nonequilibrium phase change problem. In~this~paper,
we~present a~first~study of such a dynamical transition in~magnetized QCD matter; 
to wit: we~study the nonequilibrium dynamics of the chiral condensate under a strong 
magnetic field.  

Magnetic field strengths and space-time scales in this study concern the phenomenology 
related to high-energy heavy-ion collision experiments. Relativistic heavy-ion collisions 
produce QCD matter of deconfined quarks and gluons, the quark-gluon plasma (QGP). Noncentral collisions produce the QGP under strong magnetic fields~\cite{Rafelski:1975rf,Kharzeev:2007jp}; 
for example, noncentral Pb-Pb collisions at the Large Hadron Collider (LHC) can produce fields 
of strengths as large as~\cite{Skokov:2009qp} $eB = 15 m^2_\pi$. We conduct our study within 
the perspective
of a standard three-stage scenario of QGP's time evolution~\cite{Jacak:2012dx,Shuryak:2014zxa,
Pasechnik:2016wkt,Braun-Munzinger:2015hba}: (1)~quarks and gluons are freed from 
the protons and neutrons of the colliding ions and form (2)~hot matter that expands 
hydrodynamically until it~(3)~cools up to a temperature $T \sim 150$~MeV, and finally 
disassembles into hadrons. More specifically, we work within the perspective that
local thermodynamic properties as temperature and order parameter acquire physical 
meaning. 

We address magnetic field effects on the chiral condensate dynamics with Langevin 
field equations, equations widely used in field theory treatments of dynamical phase  
transitions~\cite{{Goldenfeld:1992qy},{Onuki:2002}}. A~prototype dynamical transition 
addressed by these equations is the one of a temperature quench in a spin system, in that
a sudden drop in the system's temperature takes the system out from a spin-disordered 
phase and drives it irreversibly toward a spin-ordered phase. The quench-induced transition 
just described resembles, albeit with differences, the heavy-ion collision evolution 
across the crossover from a quark-gluon phase, in which the condensate is very small, 
toward a hadron-dominated phase, in which the condensate ultimately reaches its vacuum 
value. Indeed, for zero magnetic field, there is a vast literature on the use of Langevin field 
equations in this context{\textemdash}Refs.~\cite{Rajagopal:1993ah,Bedaque:1993fa,Greiner:1996dx,
Biro:1997va,Rischke:1998qy,Xu:1999aq,Fraga:2004hp,Boyanovsky:2006bf,Farias:2007xc,Nahrgang:2011mg,
Nahrgang:2011mv,Nahrgang:2011vn,Singh2011176,Krein:2012zt,Cassol-Seewald:2012tcq,Singh201312,
Herold:2013bi,Singh:2013pxa,Bluhm:2018qkf,Wu:2018twy} are a sample of this literature. The Langevin 
equations featured in that literature are either postulated on phenomenological 
grounds~\cite{Biro:1997va,Fraga:2004hp,Singh201312,Krein:2012zt,Cassol-Seewald:2012tcq,Wu:2018twy}, 
or derived from a microscopic model through a coarse-graining procedure~\cite{Rajagopal:1993ah,
Bedaque:1993fa,Greiner:1996dx,Rischke:1998qy,Xu:1999aq,Boyanovsky:2006bf,Farias:2007xc,Nahrgang:2011mg,
Nahrgang:2011mv,Nahrgang:2011vn,Singh2011176,Herold:2013bi,Singh:2013pxa,Bluhm:2018qkf}. We follow the 
latter approach.

We extend the semiclassical approach of Ref.~\cite{Nahrgang:2011mg} to include 
magnetic field effects on the chiral condensate dynamics. In that approach,
the~condensate dynamics is governed by a Langevin field equation derived from a semiclassical 
two-particle irreducible (2PI) effective action. The~effective action, computed with the time 
path formalism of nonequilibrium quantum field theory~\cite{Calzetta:2008iqa,Bellac:2011kqa}, 
refers to the Gell-Mann--Levy linear sigma model~\cite{GellMann:1960np} with quarks (LSMq). 
The~LSMq features degrees of freedom associated with the long-wavelength QCD 
chiral physics: constituent quarks, pseudoscalar-isoscalar mesons (pions, pseudo-Goldstone bosons), 
and a scalar-isoscalar meson (the quark condensate). The model does not describe quark 
confinement. 
Despite of this limitation, the model describes many of the equilibrium, time-independent 
magnetic field effects on the QCD equation of state, phase structure and chiral 
condensate~\cite{Fraga:2008qn,Ayala:2009ji,Frasca:2011zn,Andersen:2011ip,Andersen:2012bq,
Ruggieri:2013cya,Fraga:2013ova,
Kamikado:2013pya,Ruggieri:2014bqa,Ayala:2014mla,Ayala:2014gwa,Andersen:2014oaa,Ayala:2015lta} 
brought out by lattice calculations. We direct the reader to Refs.~\cite{Gatto:2012sp,Ayala:2014yla,
Miransky:2015ava,Andersen:2014xxa} for reviews with additional references on works 
employing the LSMq and also other models. 

This first study aims primarily to get insight into how a strong magnetic field affects 
condensate dynamics. To fulfil this aim, we simplify the analysis by omitting physical 
effects peculiar to a heavy-ion collision. We address the omissions and ensuing
consequences in the course of the presentation of our work. Besides, we seek an analytical 
understanding and avoid, whenever possible numerical calculations.  Notwithstanding the 
simplifications, our study brings new insight into a complex problem that offers enormous 
opportunities to learn about QCD matter. 

We organize the presentation of the paper as follows. In the next section, we define 
the chiral quark model upon which we base our study and summarize its main features. 
In Section~\ref{sec:langevin} we define the effective action and use the closed time path 
formalism to derive an equation of motion for the condensate, a Langevin equation
featuring dissipation and noise kernels. The latter require the magnetized thermal 
quark propagator in the real time formalism. We derive the propagator 
in Section~\ref{sec:propagator}. We complete the calculation of the the damping and
noise kernels in Section~\ref{sec:rhoDN}. We present explicit numerical results in
Section~\ref{sec:results} and conclude in Section~\ref{sec:conclusions}.

%
\section{The model}
\label{sec:model}

We present the main ingredients of the model upon which we base our study of magnetic field 
effects on the chiral condensate dynamics. The~condensate dynamics is governed by a Langevin field 
equation derived from a semiclassical two-particle irreducible (2PI) effective action~\cite{Nahrgang:2011mg}.
The~effective action builds on effective degrees of freedom associated with the long wavelength chiral physics
described by a Lagrangian featuring the approximate $SU(2)_L\times SU(2)_R$ symmetry of QCD. The~Lagrangian 
is that of the Gell-Mann--Levy linear sigma model~\cite{GellMann:1960np}, in which quarks replace 
the nucleons of the original model. As in the Lagrangian of the original model, a fermion 
isodoublet field, $q = (u,d)^T$, representing the light $u$ and $d$ quarks, Yukawa-couples to  
pseudoscalar-isotriplet pion~$\bpi$~field and a scalar-isoscalar $\sigma$~field. 
The~Lagrangian density of the linear sigma model with quarks (LSMq) is given by

\begin{equation}
{\mathcal L} = \bar{q}[i \slashed{\partial} 
- g(\sigma + i\gamma_5 \btau \cdot \bpi ) ] q 
+ \frac{1}{2}\left[ \partial_{\mu}\sigma\partial^{\mu}\sigma 
+ \partial_{\mu}\bpi\cdot\partial^{\mu}\bpi\right] - U (\sigma,\bpi) ,
\label{Lag-B0}
\end{equation}

\noindent
where $U(\sigma,\bpi)$ is the potential 

\begin{equation}
U(\sigma,\bpi) = \frac{\lambda}{4} (\sigma^2 + \bpi^2 -v^2)^2 - h_{q}\sigma - U_0,
\label{U}
\end{equation}

\noindent
where $U_0$ is an arbitrary constant setting the zero of $U(\sigma,\bpi)$.
We use the metric signature $g^{\mu\nu} = (1, -1, -1, -1)$ and the Bjorken-Drell~\cite{Bjorken:1965zz}
conventions for the Dirac $\gamma^{\mu}$ matrices, for which $\{\gamma^\mu, \gamma^\nu\} = 2 g^{\mu\nu}$.

For $h_q = 0$, the Lagrangian density is invariant under chiral $SU(2)_L\times SU(2)_R$
transformations. This symmetry can break spontaneously, in that $\sigma$~acquires a nonzero vacuum 
expectation value $\langle \sigma \rangle = v \neq 0$, whereas~$\langle \bpi \rangle = 0$ 
due to parity. 
For $h_q \neq 0$, the term ${\cal L}^{\rm LSM}_{SB} = h_q \sigma$ breaks the 
symmetry explicitly and plays the role of the symmetry-breaking quark mass term 
in the QCD Lagrangian, 
${\cal L}^{\rm QCD}_m = - m \bar{q}q$. Equality between the (vacuum or thermal) 
expectation values of ${\cal L}^{\rm LSM}_{SB}$ and ${\cal L}^{\rm QCD}_m$ implies
$ m \, \langle \bar{q}q\rangle_{\rm QCD} = - h_q \langle \sigma\rangle$, and 
establishes the physical correspondence between $\langle \sigma \rangle$ and the quark 
condensate in~QCD{\textemdash}Ref.~\cite{Koch:1997ei} presents a didactic review on this 
and other topics relating the LSM and QCD.
One can fit the parameters of the
model to chiral physics observables{\textemdash}a fit at the classical level, for example, sets 
the parameters as: $h_q = f_\pi m^2_\pi$, $v^2 = f^2_\pi - m^2_\pi/\lambda^2$, 
$m^2_\sigma = 2  \lambda^2 f^2_\pi + m^2_\pi$, and $m_q = g \langle \sigma \rangle$. 
Here $f_\pi$ and $m_\pi$ are the pion weak-decay constant and mass, $m_\sigma$ the 
$\sigma$-meson mass, and $m_q$ the constituent quark mass. 
We chose $U_0$ such that $U(0, 0) = 0$ (Ref. [29] chooses $U_0$ 
such that $U(f_\pi, 0) = 0$).

The parameter $g$ plays a very important role in the model's equilibrium thermodynamics.
For~example, when solving the model in the mean field approximation for zero 
baryon chemical potential, one obtains a first order transition at a 
temperature $T \simeq 123$~MeV with $g=5.5$, a second order transition at
$T \simeq 140$~MeV with $g=3.63$, and a crossover at $T \simeq 150$~MeV with 
$g=3.3$. We~restrict our study of the condensate dynamics to the situation of a 
crossover, the situation seemingly relevant for QCD. The model the has also been 
used to study equilibrium, time-independent 
magnetic field effects on the QCD equation of state, phase structure and chiral 
condensate{\textemdash}for references, we direct the reader to 
Refs.~\cite{Fraga:2008qn,Ayala:2009ji,Frasca:2011zn,Andersen:2011ip,Andersen:2012bq,
Ruggieri:2013cya,Fraga:2013ova,Kamikado:2013pya,Ruggieri:2014bqa,Ayala:2014mla,
Ayala:2014gwa,Andersen:2014oaa,Ayala:2015lta} and the reviews in 
Refs.~\cite{Gatto:2012sp,Ayala:2014yla,Miransky:2015ava,Andersen:2014xxa}.

We derive the LSMq effective action within the semiclassical framework developed for zero magnetic 
field in~Ref.~\cite{Nahrgang:2011mg}. In that framework, the long wavelength (soft) modes control the
$\sigma$ field dynamics, with the quarks providing a heat bath. In the present case, this means 
that the quarks are in equilibrium at some local temperature and local magnetic field. The 
magnetic field enters the LSMq Lagrangian by replacing in Eq.~(\ref{Lag-B0}) $\partial_\mu$ 
by $D_\mu = \partial_\mu + i q A_\mu$, where $q$ stands for the (quark or pion) electric 
charge and $A_\mu$ the electromagnetic vector field. We neglect pion fields in this first study 
but discuss in Section~\ref{sec:results} their possible implications on our results. In this 
semiclassical framework, the magnetic field is a background field, not a dynamical degree of 
freedom. The effective action is then a functional of the $\sigma(x)$ mean field and of the 
magnetic-field dependent quark propagator~$S(x,y)$. We denote the effective action by 
$\Gamma[\sigma,S]$.


\section{The effective action and Langevin equation}
\label{sec:langevin}

We summarize the main steps in the derivation of the Langevin equation for the $\sigma$ mean field 
from an effective action using the closed time path (CTP) formalism~\cite{Calzetta:2008iqa,Bellac:2011kqa}. 
In the CTP formalism, one evolves the fields over the Schwinger-Keldysh contour, an oriented 
time path ${\cal C} = {\cal C}_+ \cup {\cal C}_- $, in that the time variable~$t$ runs from an 
initial time $-\tau$ to a time $\tau$ along ${\cal C}_+$ and going back to $-\tau$ along 
${\cal C}_-$. One identifies fields on ${\cal C}_+$ with an~index~$+$, whereas those on 
${\cal C}_-$ with~$-$, i.e. $\sigma^a(x)$ and $S^{ab}(x,y)$ with $a=\pm$. A time instant 
on $C_-$ is posterior to any time instant on $C_+$. The fields on ${\cal C}_+$ and those 
on ${\cal C}_-$ are not independent fields; they couple through a 
{\em CTP boundary condition} in that they coincide at large~$\tau$ for all values of the spatial 
coordinate~\cite{Calzetta:2008iqa}. To set notation and make the paper self-contained, we mention 
that we set the speed of light $c$, the reduced Planck constant $\hbar = h/2\pi$, and the
Boltzmann constant $k_B$ to unity, and define $S^{ab}(x,y)$ as 

\bea
S^{++}(x,y) &=& \langle q(x) \overline{q}(y) \rangle \, \theta(x^0 - y^0) 
- \langle \overline{q}(y) q(x) \rangle \, \theta(y^0 - x^0),
\label{S++} \\[0.2true cm]
S^{--}(x,y) &=& \langle q(x) \overline{q}(y) \rangle \, \theta(y^0 - x^0) 
- \langle \overline{q}(y) q(x) \rangle \, \theta(x^0 - y^0) , 
\label{S--} \\[0.2true cm]
S^{+-}(x,y) &=& S^<(x,y) = - \langle \overline{q}(y) q(x) \rangle,  
\label{S+-} \\[0.2true cm]
S^{-+}(x,y) &=& S^>(x,y) = \langle q(x) \overline{q}(y) \rangle,  
\label{S-+}
\eea 

\noindent
where $\langle \cdots \rangle$ stands for averaging with respect to a density matrix specifying
the initial state. The~propagator~$S^{++}(x,y)$ is nothing else the causal Feynman propagator
and $S^{--}(x,y)$ the~corresponding anti-causal propagator; from 
the above definitions, one has:

\bea 
S^{++}(x,y) &=& S^{-+}(x,y)\,\theta(x^0 - y^0) + S^{+-}(x,y)\,\theta(y^0 - x^0), 
\label{S++rec} \\[0.2true cm]
S^{--}(x,y) &=& S^{+-}(x,y)\,\theta(x^0 - y^0) + S^{-+}(x,y)\,\theta(y^0 - x^0). 
\label{S--rec}
\eea 

The semiclassical action is given by

\beq
\Gamma[\sigma,S] = \Gamma_{\rm cl}[\sigma] + i \, {\rm Tr} \ln S - i \, {\rm Tr} \left(i\slashed{D}
- m_0\right)S + \Gamma_2[\sigma, S],
\label{Gamma}
\eeq

\noindent
where $\Gamma_{\rm cl}$ is the classical action, $m_0 = g \sigma_0$, and $\Gamma_2[\sigma,S]$ 
contains the sum of 2PI diagrams. Here, ${\rm Tr}$ stands for a spatial integration over the
Schwinger-Keldysh contour and sums over Dirac, color and flavor indices. Although one deals 
with two fields, $\sigma^+$ and $\sigma^-$, as mentioned above they are not independent, there 
is a single mean field $\sigma(x)$, and a single equation of motion~\cite{Calzetta:2008iqa}:

\beq 
\frac{\delta \Gamma[\sigma,S]}{\delta \sigma^+(x)}{\Biggl\vert}_{\sigma^-=\sigma^+=\sigma} 
= - \frac{\delta \Gamma[\sigma,S]}{\delta \sigma^-(x)}{\Biggl\vert}_{\sigma^+=\sigma^-=\sigma} = 0.
\label{var-sig}
\eeq 

\noindent
We need also the equation of motion for $S^{ab}(x,y)$: 

\beq 
\frac{\delta \Gamma[\sigma^+,\sigma^-,S]}{\delta S^{ab}(x,y)} = 0,
\label{var-S}
\eeq 

\noindent
or, equivalently: 

\beq
\left(i\sla{D} - m_0\right) S^{ab} (x,y) 
- \int_{\cal C} d^4z \, \frac{\delta \Gamma_2[\sigma,S]}{\delta S^{ac}(x,z)} \, S^{cb}(z,y) 
=  i \delta^{ab} \delta^{(4)}(x-y),
\label{SD}
\eeq 

\noindent
here ${\cal C}$ indicates that the integration runs over the Schwinger-Keldysh contour. Only one 2PI 
diagram contributes to $\Gamma_2[\sigma,S]$, a single one-loop diagram that involves the trace over 
the magnetic field dependent quark propagator, namely:

\beq 
\Gamma_2[\sigma,S] = g \int_{\cal C} d^4x \, {\rm tr}_{CDF} \left[ S^{++}(x,x) \sigma^+(x) 
+ S^{--}(x,x) \sigma^-(x) \right],  
\label{Gamma-sS}
\eeq 

\noindent
where ${\rm tr}_{Dcf}$ indicates trace over Dirac, color and flavor indices. 

Replacing Eq.~(\ref{Gamma-sS}) into Eqs.~(\ref{Gamma}) and (\ref{SD}), the last two terms in 
Eq.~(\ref{Gamma}) cancel; but  to complete the derivation of $\Gamma[\sigma,S]$, one still 
needs to solve Eq.~(\ref{SD}) for $S^{ab}$. However, to solve Eq.~(\ref{SD}) for $S^{ab}$ is 
not an easy task, even for the zero magnetic field case due to the spatiotemporal dependence 
of~$\sigma(x)$. Fortunately, the problem with magnetic field 
is still tractable within the spirit of the semiclassical approach we use here. Specifically, by 
assuming that long wavelength modes dominate the $\sigma(x)$ dynamics~\cite{Nahrgang:2011mg}, 
in that dynamical fluctuations $\delta \sigma$ build on a $\sigma_0$ background mean field, 
with $\sigma_0$ governed by a locally equilibrated quark heath bath described by a thermomagnetic quark propagator~$S_{\rm thm}$. The propagator $S_{\rm thm}$ depends on a local temperature 
and magnetic field, quantities that also drive a spatiotemporal dependence for the $\sigma_0$ 
mean field. In practice, this amounts to~split~$\sigma^a(x)$ as~follows:

\beq 
\sigma^a(x) =  \sigma^a_0(x) + \delta\sigma^a(x),
\label{split-sigma}
\eeq

\noindent
and write $S^{ab}$ as a functional power series in $\delta \sigma^a(x)$, with $S^{ab}_{\rm thm}$ 
the zeroth order term:

\beq 
S^{ab}(x,y) = S^{ab}_{\rm thm}(x,y) + \delta S^{ab}(x,y) + \delta^2 S^{ab}(x,y)
+ \cdots .
\label{split-S}
\eeq 

\noindent
When one replaces these expansions into Eq.~(\ref{SD}) and takes into account Eq.~(\ref{Gamma-sS}),
one determines $\delta S^{ab}(x,y)$, $\delta^2 S^{ab}(x,y), \cdots$ recursively. 
Specifically, the zeroth order propagator $S^{ab}_{\rm thm}$ obeys the equation 

\beq
\left[ i\slashed{D} - m_0 - g \, \sigma_0(x) \right] S^{ab}_{\rm thm}(x,y) 
= - i \delta^{ab} \delta^{(4)}(x-y),
\label{Sth-sigma}
\eeq 

\noindent
whereas the fluctuating contributions, up to the second order in $\delta\sigma$, read:   

\bea 
\delta S^{ab}(x,y) &=& - ig \int_{SK}d^4z \, S^{ac}_{\rm thm}(x,z) \delta\sigma^c(z) 
S^{cb}_{\rm thm}(z,y),
\label{delta1S}\\[0.3true cm]
\delta^2 S^{ab}(x,y) &=& - g^2 \int_{SK} d^4z d^4z'\, S^{ac}_{\rm thm}(x,z) 
\delta\sigma^c(z) S^{cd}_{\rm thm}(z,z') \delta\sigma^d(z') S^{db}_{\rm thm}(z',y) .
\label{delta2S}
\eea 

\noindent
Equation~(\ref{Sth-sigma}) evinces the role played by the $\sigma_0(x)$ background field, 
it gives quarks a local effective mass $m_q(x) = g\sigma_0(x)$ determined by local 
temperature and magnetic field. To obtain the equation of motion 
for the mean field, one can now replace Eqs.~(\ref{split-sigma})-(\ref{delta2S}) 
into Eq.~(\ref{Gamma}) and trail the steps in Ref.~\cite{Nahrgang:2011mg}. Although the 
magnetic field introduces new features into the Langevin dynamics, the generic form of the 
equation is the same as for zero magnetic field, in that $S^{ab}_{\rm thm}$ contains all 
the effects of the magnetic field on the $\sigma$ dynamics. Therefore, for now, we 
do not need the explicit expression for $S^{ab}_{\rm thm}$ to write down the Langevin 
equation{\textemdash}we obtain the explicit form of $S^{ab}_{\rm thm}$ in the following 
section. 

But before writing down the Langevin equation for the $\sigma$ mean field, we comment on 
two points in 
the derivation of the equation, namely: the lack of independence of the fields on ${\cal C}_+$ 
from those on ${\cal C}_-$, and the appearance of the noise source in the $\delta\sigma$ equation 
of motion. To~account for the first point, one performs a change of basis~\cite{Calzetta:2008iqa}, 
a.k.a. Keldysh rotation~\cite{Kamenev:2011}. We apply the Keldysh rotation to~$\sigma = \sigma_0 
+\delta \sigma$, which implies for the fluctuating $\delta\sigma$ field needed here:   

\beq 
\delta\bar\sigma(x) =  \frac{1}{2} \left(\delta\sigma^+(x) + \delta\sigma^-(x) \right), 
\hspace{1.0cm}
\Delta\sigma(x) = \delta\sigma^+(x) - \delta\sigma^-(x) .
\label{sigma-Kel}
\eeq

\noindent
This transformation makes transparent the physics behind the doubling of fields: it
reflects the need for both {\em response} ($\Delta\sigma$) and {\em fluctuating} 
($\delta\bar\sigma$) fields to describe time-dependent fluctuating phenomena~\cite{Martin:1973zz}. 
The second point refers to the fact that $\Gamma[\sigma,S]$ contains an imaginary part associated
with dissipation, a feature that obstructs the straightforward variation implied by 
Eq.~(\ref{var-sig}). A~way to~obtain a~real action uses the Feynman-Vernon 
trick~\cite{Feynman:1963fq}, in that one replaces the imaginary part of 
the action by a noise source coupling linearly to the field; this turns the equation 
of motion into a 
stochastic equation{\textemdash}we refer to the book of Ref.~\cite{Calzetta:2008iqa} 
for a thorough discussion on this and other aspects of the CTP formalism. In summary, after using 
Eqs.~(\ref{split-sigma})-(\ref{delta2S}) into Eq.~(\ref{Gamma}) and rewriting the action in terms of
the Keldysh-rotated fields, replacing the resulting imaginary part in the action by a noise source, 
and varying w.r.t. $\Delta\sigma$ and setting $\bar\sigma(x) = \sigma(x)$ as implied by 
Eq.~(\ref{var-sig}), one obtains a stochastic differential equation for $\sigma(x)$, 
namely~\cite{Nahrgang:2011mg}:

\beq 
\partial_\mu\partial^\mu \sigma(x) + \frac{\delta U[\sigma]}{\delta \sigma(x)} + g \rho_s(\sigma_0)
- D_{\sigma}(x) = \xi_{\sigma}(x) ,
\label{EoM}
\eeq 

\noindent
where $\rho_s(\sigma_0)$ is the scalar density:

\beq 
\rho_s(\sigma_0) =  \mathrm{tr}_{Dcf} \, S_{thm}^{++}(x,x) ,
\label{rhos}
\eeq 

\noindent
and $D_{\sigma}(x)$ the {\em dissipation kernel}: 

\beq 
D_{\sigma}(x) = i g^2 \int d^4 y \, \theta(x^0 - y^0) \, M(x,y) \, \delta\bar\sigma(y),
\label{D}
\eeq 

\noindent
with 

\beq 
M(x,y) =  \mathrm{tr}_{Dcf}
\left[ 
S^{+-}_{thm}(x,y) S^{-+}_{thm}(y,x) - S^{-+}_{thm}(x,y)S^{+-}_{thm}(y,x)
\right],
\label{M}
\eeq 

\noindent
and $\xi_{\sigma}(x)$ is a colored-noise field with the properties:

\beq 
\langle \xi_\sigma(x) \rangle_\xi = 0, \hspace{1.0cm}
\langle \xi_\sigma(x) \xi_\sigma(y) \rangle_\xi = N(x,y),
\label{noise-corr}
\eeq 

\noindent
with the {\em noise kernel} $N(x,y)$ given by: 

\beq 
N(x,y) =  - \frac{1}{2} g^2 \mathrm{tr}_{Dcf}
\left[ S^{+-}_{thm}(x,y) S^{-+}_{thm}(y,x) +  S^{-+}_{thm}(x,y) S^{+-}_{thm}(y,x) \right].
\label{NoiseKernel}
\eeq 

\noindent
In Eq.~(\ref{noise-corr}), $\langle \cdots \rangle_\xi$ means functional average 
with the probability distribution 

\beq
P[\xi] = \exp \left[-\frac{1}{2} \int d^4x d^4y \, \xi(x) N^{-1}(x,y) \, \xi(y) \right].
\label{prob-dist}
\eeq 

Our summary on the the derivation of the Langevin equation ends here. To proceed with the study 
of magnetic field effects on the $\sigma$ dynamics, we need the explicit form of the thermomagnetic 
quark propagator $S^{ab}_{thm}${\textemdash}we derive $S^{ab}_{thm}$ in the next 
section. 


\section{The thermomagnetic quark propagator}
\label{sec:propagator}

One can obtain the CTP thermomagnetic quark propagator at temperature~$T$ from the corresponding 
$T=0$ propagator through a Bogoliubov transformation, much in the same way as done in thermofield 
dynamics (TFD)~\cite{Loewe:1991mn,Elmfors:1993wj,Hasan:2017fmf,Rath:2017fdv}. 
Let $\overline{S}_m(x,y)$ 
be the causal, zero temperature quark propagator in a  constant magnetic field of strength~$B$ 
pointing along the $\hat{\bz}$ direction. $\overline{S}_m(x,y)$ can be written as the product of a 
gauge-dependent 
Schwinger phase $\phi(x,y)$ and a gauge-independent, translation invariant propagator 
$S_m(x-y)$, namely~\cite{Schwinger:1951nm}:

\bea 
\overline{S}_m(x,y) &=& \theta(x^0 - y^0) \langle q(x) \bar{q}(y) \rangle 
- \theta(y^0 - x^0) \rangle \bar{q}(y)\psi(y)\rangle = S^{++}_m(x,y)  
\label{S_m} \\[0.2true cm]
&=& \theta(x^0 - y^0) S^{-+}_m(x,y) + \theta(y^0-x^0) S^{+-}_m(y,x) 
\label{S+-m}  \\
&=& \phi(x,y) \, S_m(x-y) = \phi(x,y) \, \int\frac{d^4p}{(2\pi)^4} \, 
e^{-i p\cdot(x-y)}\, S^{++}_m(p) .
\label{overS}
\eea 

\noindent
The phase factor is irrelevant for us since it cancels out in all terms appearing in the
Langevin equation in Eq.~(\ref{EoM}) due to the properties $\phi(x,z) \phi(z,y) = \phi(x,y)$
and $\phi(x,x) = 1$. We~can~also choose the symmetric gauge, $A_\mu = (0, -y, x, 0) \, B/2$, for which $\phi(x,y) = 1$. In any case, one can focus on the translation invariant piece of the
propagator which, from now on will be the object of interest. 

We use the Landau level representation of $S_m(p)$ and work with the lowest level contribution, the 
dominant contribution for strong fields. The lowest Landau level (LLL) contribution to $S_m(p)$ can
be written as~\cite{Miransky:2015ava}:

\beq 
S^{++}_m(p) = i \, e^{-{\bp^2_\perp}/{\vert q_f B\vert}} \,
\frac{2\left(\slashed{p}_{\parallel} + m_q\right)}{p_{\parallel}^2 - m^2_q + i\epsilon} \, P_+,
\label{S++m}
\eeq 

\noindent
where $\bp^2_\perp = p^2_x + p^2_y$, $p_{\parallel}^2 = p_{0}^2 -p_{z}^2$, 
$\slashed{p}_{\parallel} = \gamma^0 p_0 - \gamma^3 p_z$, and 
$P_+ = [1 + i \gamma^1 \gamma^2 {\rm sign}(qB)]/2$. The presence of the operator $P_+$ in 
Eq.~(\ref{S++m}) reflects the spin-polarized nature of the lowest Landau level, as
$P_+$ projects out 
one of the two spin directions. From this result, one obtains the off-diagonal CTP components
$S^{+-}_m$ and $S^{-+}_m$ by using Eqs.~(\ref{S++rec}) and (\ref{S+-m}) and the identity:

\bea
i \int^\infty_{-\infty} dp_0  \frac{ \slashed{p}_\parallel + m_q}{p^2_\parallel - m^2_q 
+ i \epsilon} \, 
e^{- i p_0 (x^0 - y^0)}
&=& \theta(x^0 - y^0) \int^\infty_{-\infty} dp_0 
(\slashed{p}_\parallel + m_q) 2\pi \delta(p^2_\parallel - m^2_q)\theta(p_0) \, 
e^{-i p_0 (x^0 - y^0)} 
\nn \\[0.1true cm]
&&\hspace{-1.5cm}  + \, \theta(y^0 - x^0) \int^\infty_{-\infty} dp_0 (\slashed{p}_\parallel + m_q)
2\pi \delta(p^2_\parallel- m^2_q)\theta(-p_0) e^{- i p_0 (x^0 - y^0)}. 
\eea 

\noindent
Therefore, the CTP components $S^{ab}_m(p)$ of the zero-temperature propagator can be written as: 

\bea 
S^{++}_m(p) &=&  e^{-{\bp^2_\perp}/{\vert q_f B\vert}} \, A(p) \, 
\frac{i}{p^2_\parallel - m^2_q + i\epsilon}, 
\label{S++m-p} \\[0.2true cm]
S^{+-}_m(p) &=& e^{-{\bp^2_\perp}/{\vert q_f B\vert}} \, A(p)  \, 
2\pi \delta(p^2_\parallel - m^2_q) \, \theta(-p_0) ,
\label{S+-m-p} \\[0.2true cm]
S^{-+}_m(p) &=& e^{-{\bp^2_\perp}/{\vert q_f B\vert}} \, A(p) \, 
2\pi \delta(p^2_\parallel - m^2_q) \, \theta(p_0) ,
\label{S-+m-p} \\[0.2true cm]
S^{--}_m(p) &=&  e^{-{\bp^2_\perp}/{\vert q_f B\vert}} \, A(p) \, 
\frac{-i}{p^2_\parallel - m^2_q - i\epsilon}, 
\label{S--m-p}
\eea 

\noindent
where, to lighten the notation, we defined

\beq
A(p) = 2 (\slashed{p}_{\parallel} + m_q) \, P_+ = (\slashed{p}_{\parallel} + m_q)
\left[1 + i \gamma^1 \gamma^2 {\rm sign}(qB) \right] .
\label{A-def}
\eeq 

One obtains the thermal propagator $S^{ab}_{thm}(p)$ from $S^{ab}_m(p)$ through the
Bogoliubov transformation,

\beq 
\begin{pmatrix}
S^{++}_{thm}(p) &  S^{+-}_{thm}(p) \\[0.2true cm]
S^{-+}_{thm}(p) & S^{--}_{thm}(p) 
\end{pmatrix}
= V_{CTP}(T,p) 
\begin{pmatrix}
S^{++}_{m}(p) &  S^{+-}_{m}(p) \\[0.2true cm]
S^{-+}_{m}(p) & S^{--}_{m}(p) 
\end{pmatrix}
V_{CTP}(T,p) .
\eeq 

\noindent
A possible CTP transformation matrix $V_{CTP}(T,p)$ is the following: 

\beq 
V_{CTP}(T,p) = \frac{1}{2\sqrt{\sinh|p_0|/T}}
\begin{pmatrix} 
e^{|p_0|/2T}    & - e^{-|p_0|/2T}  \\[0.2true cm]
- e^{-|p_0|/2T} &   e^{|p_0|/2T}
\end{pmatrix} .
\eeq 

\noindent
$V_{CTP}$ is the fermionic counterpart to the bosonic Bogoliubov transformation 
matrix in Ref.~\cite{Das:2018qak}, denoted $U_{CT}(T,p)$ in that reference. 
The individual $S^{ab}_{thm}(p)$ components are then given by: 

\bea 
S^{++}_{thm}(p) &=& e^{-{\bp^2_\perp}/{\vert q_f B\vert}}  \, A(p) 
\left[\frac{i}{p_{\parallel}^2 - m^2_q + i\epsilon} 
- 2\pi n_F(p_0) \delta(p^2_\parallel - m^2_q) \right],
\label{S++thm} \\[0.2true cm]
S^{+-}_{thm}(p) &=& e^{-{\bp^2_\perp}/{\vert q_f B\vert}}  \, A(p) 
2\pi \delta(p^2_\parallel - m^2_q) \left[\theta(-p_0) - n_F(p_0)\right] ,
\label{S+-thm} \\[0.2true cm]
S^{-+}_{thm}(p) &=& e^{-{\bp^2_\perp}/{\vert q_f B\vert}}  \, A(p) 
2\pi \delta(p^2_\parallel - m^2_q) \left[\theta(p_0) - n_F(p_0)\right] ,
\label{S-+thm} \\[0.2true cm]
S^{--}_{thm}(p) &=& e^{-{\bp^2_\perp}/{\vert q_f B\vert}}  \, A(p) 
\left[\frac{-i}{p_{\parallel}^2 - m^2_q - i\epsilon} 
- 2\pi n_F(p_0) \delta(p^2_\parallel - m^2_q) \right],
\label{S--thm}
\eea 

\noindent
where $n_F(p_0)$ is the Fermi-Dirac distribution:

\beq 
n_F(p_0) = \frac{1}{e^{|p_0|}/T + 1}.
\label{FD}
\eeq 

\noindent
Here, $q_u = 2 e/3$, $q_d = - e/3$, and $e = 1/\sqrt{137}$ 
(we use Gaussian units).
We note that one obtains
the same result for $S^{ab}_{thm}$ with the more standard TFD Bogoliubov transformation,
by multiplying the off-diagonal elements of TFD propagator, $S^{12}(p)$ and $S^{21}(p)$,
by $e^{-p_0/2T}$ and $e^{+p_0/2T}$, respectively. The diagonal elements of CTP and TFD 
propagators are the same, of course.
 
We note that the LLL approximation is suitable for strong magnetic fields only. Therefore, one 
cannot extrapolate $B \neq 0$ results to recover $B = 0$ results. Such an extrapolation 
is possible when performing the sum over all Landau levels or using an alternative 
representation of the propagator{\textemdash}see, for example, Appendix~A of 
Ref.~\cite{Miransky:2015ava}.  

This completes the derivation of $S^{ab}(x,y)$. In the next section we compute the different
pieces entering the Langevin equation in Eq.~(\ref{EoM}), namely, the scalar density $\rho_s$
and the dissipation $D(x)$ and noise $N(x,y)$ kernels. As mentioned before, our interest in
on the long-wavelength physics of the $\sigma$ mean field dynamics, thereby we neglect vacuum 
contributions to these quantities.


\section{The scalar density, dissipation and noise kernels}
\label{sec:rhoDN}

We start with the scalar density~$\rho_s(\sigma_0)$. Although we have flavor symmetry at the level 
of the quark masses, $m_u = m_d = m_q = g \sigma_0$, we still need to make explicit the flavor 
content of the propagator because of the quark electric charges. After taking the trace over Dirac, 
color and flavor indices, one can write $\rho_s(\sigma_0)$ as the sum of two 
contributions~\cite{Menezes:2008qt,Farias:2014eca}, $\rho_s(\sigma_0) = \rho^B_s(\sigma_0) 
+ \rho^{BT}_s(\sigma_0)$, 
where $\rho^B_s(\sigma_0)$ depends only on $B$:

\beq 
\rho^B_s(\sigma_0) = - \frac{N_c}{2\pi^2} \, m_q\, \sum_{f=u,d} |q_fB| \left[\ln\Gamma(x_f) 
- \frac{1}{2} \ln 2\pi  + x_f - \frac{1}{2}(2x_f -1)\ln x_f\right],
\label{rhoB}
\eeq 

\noindent
and $\rho^{BT}_s(\sigma_0)$ that depends on $B$ and $T$:  

%

\beq 
\rho^{BT}_s(\sigma_0) = - \frac{N_c}{\pi^2} \, m_q \left(|q_u B|+|q_d B|\right)  
\int^\infty_0 dp_z \, \frac{n_F(E_q(p_z))}{E_q(p_z)} ,
%
\eeq  

\noindent
where $N_c =3$ is the number of colors, $x_f = m^2_q/2|q_fB|$, $\Gamma(x)$ the Euler gamma
function, and $E_q(p_z) = \sqrt{p^2_z + m^2_q}$. 

%
%

Next, we consider the dissipation $D(x)$ and noise $N(x,y)$ kernels , Eqs.~(\ref{D})
and (\ref{NoiseKernel}). To compute $D(x)$, we need the function $M(x,y)$, given in
Eq.~(\ref{M}). The Schwinger phase $\phi(x,y)$ cancels out in Eq.~(\ref{D}); as a result, 
$M(x,y)$ becomes a function of $x-y$:

\begin{eqnarray}
M(x - y) &=& \mathrm{tr}_{f} \int \frac{d^4p}{(2\pi)^4} \frac{d^4q}{(2\pi)^4} \,
e^{- i (p-q) \cdot(x-y)} \, \mathrm{tr}_{Dc} \left[ S^{+-}_{thm}(p) S^{-+}_{thm}(q) 
- S^{-+}_{thm}(p)S^{+-}_{thm}(q)
\right] 
\nn\\[0.2true cm]
&=& \int \frac{d^4p}{(2\pi)^4} e^{- i p \cdot(x-y)}  \, M(p) ,
\end{eqnarray}

\noindent
where

\bea 
M(p) = \sum_{f=u,d}  \int \frac{d^4q}{(2\pi)^4} \, \mathrm{tr}_{Dc} 
\left[ S^{+-}_{thm}(p+q) S^{-+}_{thm}(q)
- S^{-+}_{thm}(p+q) S^{+-}_{thm}(q) \right]_f 
\equiv  \sum_{f=u,d} M_f(p) . 
\label{M-Mf}
\eea  

\noindent
The Schwinger phase also cancels out in Eq.~(\ref{NoiseKernel}) and $N(x,y) = N(x-y)$ 
can be written as:

\bea 
N(x-y) = \int \frac{d^4p}{(2\pi)^4}\, e^{-ip\cdot(x-y)} \, N(p) ,
\eea 

\noindent
where

\beq 
N(p) =  - \frac{1}{2} g^2 \sum_{f=u,d} \int \frac{d^4q}{(2\pi)^4} \, \mathrm{tr}_{Dc} 
\left[ S^{+-}_{thm}(p+q) S^{-+}_{thm}(q)
+  S^{-+}_{thm}(p+q) S^{+-}_{thm}(q) \right]_f \equiv \sum_{f=u,d}
N_f(p). 
\eeq

Next, we use $M$'s translation invariance to write the dissipation kernel $D(x)$ 
as~\cite{Nahrgang:2011mg}:

\bea 
D(x) &=& D(t,\bx) = i g^2 \int \frac{d^4p}{(2\pi)^4} \, M(p_0,\bp) 
\int d^3y dy^0 \, \theta(x^0 - y^0) 
e^{-i p_0 (x^0 - y^0) + i \bp \cdot (\bx - \by)} \, 
\delta\sigma(y^0,\by)
\nn \\[0.3true cm]
&=& i g^2 \int \frac{d^3p}{(2\pi)^3} \, e^{ i \bp \cdot \bx} 
\int^\infty_{-\infty} \frac{dp_0}{2\pi} \, M(p_0,\bp)\, 
\int^\infty_0 d\tau \, e^{- i p_0 \tau} \, 
\delta\sigma(t-\tau,\bp).
\label{D-mem}
\eea 

\noindent
Here, we made the change of variable $x^0 - y^0 = \tau$ and 
defined the spatial Fourier transform of the $\sigma$ mean field:

\beq 
\delta\sigma(t-\tau,\bp) = \int d^3y \, e^{-i\bp\cdot\by}\, 
\delta\sigma(t-\tau,\by).
\label{FT-sig}
\eeq 

\noindent
Eq.~(\ref{D-mem}) exposes the presence of memory in the $\sigma$ dynamics, 
in that the value of $\sigma$ at time $t$ depends upon the values of $\sigma$ at earlier 
times $t-\tau$. This feature imposes technical difficulties to the analysis of the Langevin
equation as it requires numerical techniques to proceed. To maintain the pace with an analytically 
tractable analysis, we follow Refs.~\cite{Greiner:1996dx,Rischke:1998qy,Nahrgang:2011mg} and use 
a {\em linear harmonic approximation}, whereby the dynamics memory is captured by soft-mode 
harmonic oscillations around the mean field $\sigma_0(t,\bp)$. This approximation amounts to assume 
an harmonic $\tau$ dependence for $\bar\sigma(t-\tau,\bp)$, namely:

\bea
\sigma(t-\tau,\bp) &=& a(t) \cos(E_\sigma(\bp) \tau) + b(t) \sin (E_\sigma(\bp)\tau) 
\nn \\[0.2true cm]
&=& \sigma_0(t,\bp) \cos(E_\sigma(\bp)\tau) - \frac{1}{E_\sigma(\bp)} \, \sin (E_\sigma(\bp) \tau)\,
\frac{\partial\sigma(t,\bp)}{\partial t} 
\nn \\[0.2true cm]
&\equiv& \sigma_0(t,\bp) + \delta\sigma(t,\bp) \nn \\[0.2true cm]
&=& 
\sigma_0(t,\bp) + \left\{ 
\sigma_0(t,\bp) [\cos(E_\sigma(\bp)\tau) - 1 ]
- \frac{1}{E_\sigma(\bp)} \, \sin (E_\sigma(\bp) \tau)\,
\frac{\partial\sigma(t,\bp)}{\partial t} \right\} ,
\label{harmonic}
\eea 

\noindent
where $E_\sigma(\bp) \approx \sqrt{\bp^2 + m^2_\sigma}$ is a characteristic soft-mode 
frequency, where $m_\sigma$ is the $\sigma$ field mass. The functions $a(t)$ and $b(t)$ were 
determined using as initial conditions $\sigma(t-\tau,\bp)|_{\tau=0} = \sigma_0(t,\bp)$ and 
$\partial \sigma(t-\tau,\bp)/\partial\tau|_{\tau=0} = - \partial \sigma(t,\bp)/\partial t$. 
The first term within the curly brackets in Eq.~(\ref{harmonic}), being linear in $\sigma_0$ is a 
leading-order correction to $ g \rho_s$ and, since $\cos(E_\sigma(\bp)\tau) - 1$ oscillates 
around zero, it is neglected; as such, one obtains for $D(t,\bx)$:

\bea  
D(t,\bx) &=& -  \int \frac{d^3p}{(2\pi)^3} \, e^{i\bp\cdot\bx} \, \eta(\bp) \,
\frac{\partial \sigma(t,\bp)}{\partial t}. 
\label{D-local}
\eea 

\noindent
where $\eta(\bp)$ is the momentum-dependent {\em damping coefficient}:

\beq 
\eta(\bp) = g^2 \frac{1}{2E_\sigma(\bp)} M(\bp) .
\label{damping-coeff} 
\eeq 

\noindent
To lighten the notation, we denoted $\eta(E_\sigma(\bp),\bp)$ by $\eta(\bp)$ and 
$M(E_\sigma(\bp), \bp)$ 
by $M(\bp)${\textemdash}from this point on, this notation will be used throughout the paper. 

The harmonic approximation rendered the dissipation kernel local in time and
in a form appropriate to work with the Langevin equation in momentum-space:

\beq 
\frac{\partial^2\sigma(t,\bp)}{\partial t^2} 
+ \bp^2 \, \sigma(t,\bp) + \eta (\bp) \,
\frac{\partial \sigma(t,\bp)}{\partial t} 
+ F_\sigma(t,\bp) 
= \xi_{\sigma}(t,\bp), 
\label{EoM-mom}
\eeq 

\noindent
where $\eta(\bp)$ was defined in Eq.~(\ref{damping-coeff}), and 

\bea 
F_\sigma(t,\bp) = \int d^3x \, e^{-i \bp \cdot \bx} \, \left[
\frac{\delta U[\sigma]}{\delta \sigma(t,\bx)} + g\,\rho_s(\sigma_0) \right].
\label{def-f}
\eea 

\noindent
The momentum space colored noise field has zero mean $\langle \xi_{\sigma}(t,\bp) \rangle_\xi = 0$
and correlation:

\beq 
\langle \xi_\sigma(t, \bp)  \xi_\sigma(t, \bp) \rangle_\xi = 
(2\pi)^3 \delta(\bp + \bp')  N(t-t',\bp),
\label{noise-cf}
\eeq 

\noindent
where 

\beq
N(t-t',\bp) = \int^\infty_{-\infty} \frac{dp_0}{2\pi} \, e^{- i p_0 (t - t')} \, N(p_0,\bp).
\label{N(t,p)}
\eeq 

Although Eq.~(\ref{EoM-mom}) involves colored noise, it can be solved efficiently by iteration 
on a discrete momentum lattice using fast Fourier transformation to switch back and forth between 
coordinate space and momentum space to compute the nonlinear term 
$F_\sigma(t,\bp)$~\cite{CASSOLSEEWALD2008297}. As our aim is to get analytic understanding 
as much as possible, we leave for a future publication the study 
of numerical solutions of  Eq.~(\ref{EoM-mom}).  But we need to simplify further the 
analysis to proceed 
with an analytical treatment. A~common simplification restricts the dynamics to a constant soft-mode 
frequency $E_\sigma(\bp) \approx \sqrt{\bp^2 + m^2_\sigma} \approx m_\sigma$~\cite{Rischke:1998qy,
Nahrgang:2011mv,Herold:2013bi,Bluhm:2018qkf}. We adopt another simplification, one motivated by the 
dimensional reduction brought out by the magnetic field: we restrict the dynamics to the plane 
orthogonal to the magnetic field, namely $\sigma(t, \bp) = \sigma(t, p_x, p_y, p_z) 
\rightarrow \sigma(t, p_x, p_y, p_z=0) 
\equiv \sigma(t, \bp_\perp)$. Therefore, we need to compute the kernel $M_f(\bp)$ for 
$\bp = (\bp_\perp, p_z=0)$. We use Eqs.~(\ref{S+-thm}) and (\ref{S-+thm}) into Eq.~(\ref{M-Mf}), 
take the traces over Dirac and color indices and integrate over the transverse momentum $\bq_\perp$, 
to obtain for $M_f(\bp_\perp, p_z=0) \equiv M_f(\bp_\perp) $ 
the result: 

\bea 
M_f(\bp_\perp) &=& \frac{2 N_c}{\pi^2} 
\left( \frac{\pi |q_f| B}{2} \right) 
e^{-\bp^2_\perp/2|q_fB|} \;
I_M(E_\sigma(\bp_\perp)),
\eea 

\noindent
where $I_M(E_\sigma(\bp_\perp))$ is the integral

\bea 
I_M(E_\sigma(\bp_\perp)) &=& \int^\infty_{-\infty} dq_z \int^\infty_{-\infty} dq_0 \; 
\Bigl\{ 
\delta((p+q)^2_\parallel - m^2_q) \delta(q^2_\parallel - m^2_q) 
\left[ q_0 \left(p_0 + q_0 \right) - q_z \left(p_z + q_z \right) + m^2_q \right]  \nn \\[0.2true cm]
&& \times \, \Bigl( \left[ \theta(q_0)  - n_F (q_0)\right] 
\left[ \theta(- p_0 -q_0)  - n_F (p_0 +q_0)\right]
\nn  \\[0.2true cm]
&& - \left[ \theta(-q_0)  - n_F (q_0) \right] 
\left[ \theta(p_0 +q_0)  - n_F (p_0 + q_0) \right] \Bigr)
\Bigr\}_{p_0 = E_\sigma(\bp_\perp)} .
\label{I_M}
\eea  

\noindent
Here and in the following we suppress the explicit reference to the fact that $p_z = 0$ 
in $\bp$-dependent
functions. We first use the delta function $\delta(q^2_\parallel - m^2_q) = 
\delta(q^2_0 - q^2_z - m^2_q)$ to
integrate over $q_0$, then use the other delta function to integrate over $q_z$ to obtain: 

\bea 
I_M(E_\sigma(\bp_\perp)) &=& \left[1 - 2n_F(E_\sigma(\bp_\perp)/2)\right] 
\left(\frac{E^2_\sigma(\bp_\perp) - 4 m^2_q}{4E^2_\sigma(\bp_\perp)}\right)
\int^\infty_{-\infty} dq_z  \; \delta(E_q(q_z) - E_\sigma(\bp_\perp)/2) \nn \\
&=&  \left[1 - 2n_F(E_\sigma(\bp_\perp)/2)\right] \left(\frac{E^2_\sigma(\bp_\perp) 
- 4 m^2_q}{4E^2_\sigma(\bp_\perp)}\right)
\frac{2 E_\sigma(\bp_\perp)}{\sqrt{E^2_\sigma(\bp_\perp) - 4m^2_q}} \nn \\
&=&  \left[1 - 2n_F(E_\sigma(\bp_\perp)/2)\right] \frac{1}{2E_\sigma(\bp_\perp)} 
\sqrt{E^2_\sigma(\bp_\perp) - 4m^2_q}. 
\eea 

\noindent
Therefore:
\beq 
M_f(\bp_\perp) = \frac{N_c}{\pi} 
\left(|q_f| B\right) 
\left[1 - 2n_F(E_\sigma(\bp_\perp)/2)\right] \frac{1}{2E_\sigma(\bp_\perp)} 
\sqrt{E^2_\sigma(\bp_\perp) - 4m^2_q} \; e^{-\bp^2_\perp/2|q_fB|} .
\label{M_f-fin}
\eeq 

\noindent
From this, one obtains for the momentum-dependent noise coefficient $\eta(\bp_\perp)$:

\beq 
\eta(\bp_\perp) = g^2 \frac{N_c}{4\pi} \left[1 - 2n_F(E_\sigma(\bp_\perp)/2)\right] 
\frac{1}{E^2_\sigma(\bp_\perp)}
\sqrt{E^2_\sigma(\bp_\perp) - 4m^2_q} 
\sum_{f=u,d} |q_f B|  \; e^{-\bp^2_\perp/2|q_fB|} .
\label{eta-final}
\eeq

Next, we compute the noise kernel $N(x,y)$ with the same simplifications used for $M(x)$. 
We use Eqs.~(\ref{S+-thm}) and (\ref{S-+thm}) into Eq.~(\ref{NoiseKernel}), take the traces 
over Dirac and color indices, and integrate over the transverse momentum $\bq_\perp$ 
to obtain:

\bea 
N_f(\bp_\perp) &=& - \frac{1}{2} g^2 \frac{2 N_c}{\pi^2} \left( \frac{\pi |q_f| B}{2} \right) 
e^{-\bp^2_\perp/2|q_fB|} \;
I_N(E_\sigma(\bp_\perp)),
\eea 

\noindent
with 

\beq 
I_N(E_\sigma(\bp_\perp)) = - \left[1 - 2n_F(E_\sigma(\bp_\perp)/2)\right] 
\coth (E_\sigma(\bp_\perp)/2T) \frac{1}{2E_\sigma(\bp_\perp)} 
\sqrt{E^2_\sigma(\bp_\perp) - 4m^2_q} .
\eeq 

\noindent
Taking into account Eq.~(\ref{M_f-fin}), one can write: 

\beq 
N_f(\bp_\perp)  = \frac{1}{2}\, 
g^2 \coth (E_\sigma(\bp_\perp)/2T) \, M_f(\bp_\perp).
\eeq 

\noindent
Therefore, after summing over flavor and using the result in Eq.~(\ref{eta-final}), one can
write for the momentum space noise kernel $N(\bp_\perp)$:

\beq 
N(\bp_\perp)  = \eta(\bp_\perp) \, E_\sigma(\bp_\perp) \, \coth (E_\sigma(\bp_\perp)/2T) . 
\label{noise-final}
\eeq 

\noindent
Finally, replacing this result into Eq.~(\ref{N(t,p)}), the $p_0$ integration leads to the Dirac 
delta $\delta(t - t')$ and $\xi_\sigma$ becomes a white noise field. 

This concludes the derivation of the main ingredients entering the Langevin equation: $\rho_s$, 
$D(x)$ and $N(x,y)$. In the next section, we examine the effects of a nonzero magnetic field 
on these quantities. There we also need the equilibrium mean field $\sigma_0$ and mass $m_\sigma$, 
which we discuss in the following. 

We close this section deriving the equilibrium (constant and uniform) mean field solution by putting 
to zero the time and space derivatives and the dissipation and noise kernels in the Langevin
equation in Eq.~(\ref{EoM}), so that $\sigma = \sigma_0 + \delta\sigma \rightarrow \sigma_0$ and:

\beq 
\frac{\delta U[\sigma_0]}{\delta \sigma_0} + g\rho_s(\sigma_0) = 0. 
\label{sig-equil}
\eeq 

\noindent
This equation is nothing else than the equation one obtains from the minimization of the equilibrium 
effective potential $V_{\rm eff}[\sigma_0]$:

\beq 
V_{\rm eff}[\sigma_0] = U[\sigma_0] + \Omega^{B}[\sigma_0] 
+  \Omega^{BT}[\sigma_0], 
\label{Veff}
\eeq 

\noindent
with~\cite{Ebert:1999ht,Fraga:2008qn,Menezes:2008qt} 

\bea 
\Omega^{B}[\sigma_0] &=& - \frac{N_c}{2\pi^2} \sum_{f=u,d} \left(|q_f B|\right)^2 
\left[ \zeta'(-1,x_f) - \frac{1}{2}\left(x^2_f - x_f\right) \ln x_f + \frac{1}{4} x^2_f\right], 
\label{Omega-B} \\[0.2true cm]
\Omega^{BT}[\sigma_0] &=& - \frac{N_c}{\pi^2} T \sum_{f=u,d} |q_fB|  
\int^\infty_0 dp_z\, \ln\left(1 + e^{-E_q(p_z)/T}\right)  ,
\label{Omega-BT}
\eea

\noindent
where $\zeta'(-1,x) = d\zeta(s,x)/ds|_{s=-1}$ and $\zeta(s,x)$ the Riemann-Hurwitz
zeta function. 
That is:

\beq
g \rho^B_s(\sigma_0) = \frac{\delta\Omega^B[\sigma_0]}{\delta\sigma_0} 
\hspace{0.5cm}\text{and}\hspace{0.5cm}
g  \rho^{BT}_s(\sigma_0) = \frac{\delta\Omega^{BT}[\sigma_0]}{\delta\sigma_0} .
\eeq

\noindent
We used the result $d\zeta'(-1,x)/dx = - 1/2 + x + \ln\Gamma(x) - 1/2 \ln 2\pi$ to obtain the 
expression
for $g \rho^B_s(\sigma_0)$. We obtain the temperature and magnetic field dependent mean field mass 
$m_\sigma$ from:

\beq 
m^2_\sigma = \frac{\delta^2 V_{\rm eff}[\sigma]}{\delta \sigma^2}\Bigg\vert_{\sigma_{\rm min}}.
\label{sig-mass}
\eeq 

\begin{figure}[h]
\begin{center}
\begin{tabular}{lr}
\includegraphics[clip,width=0.48\linewidth]{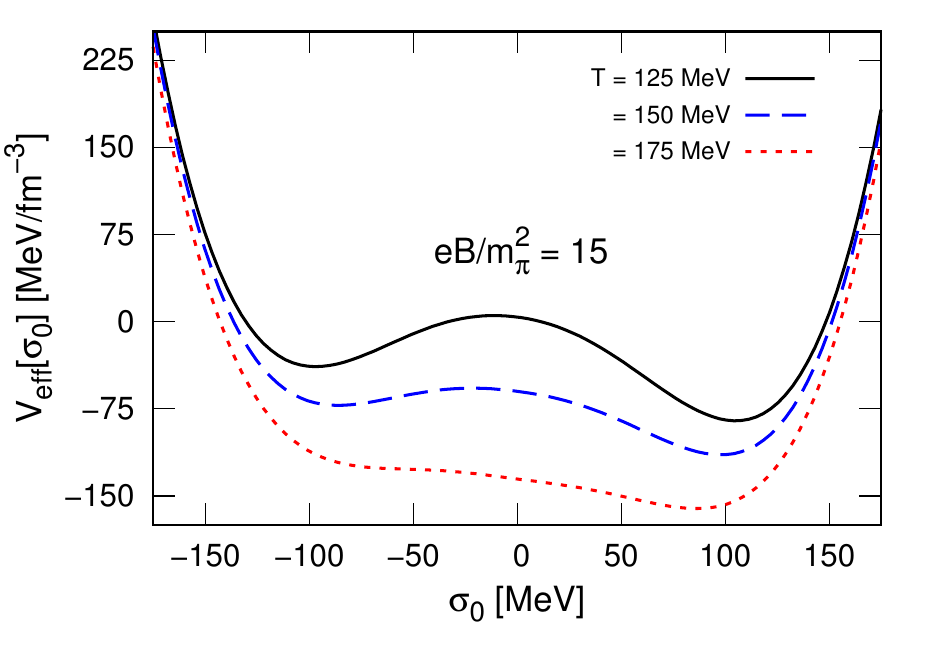} \hspace*{-2.5ex } &
\includegraphics[clip,width=0.48\linewidth]{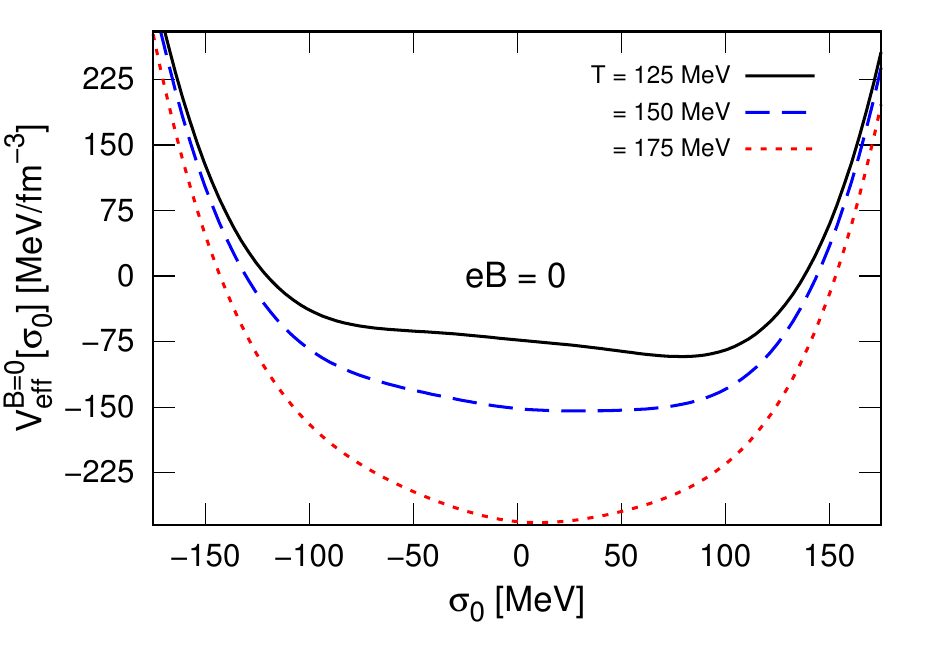}\vspace*{-0ex}  
\end{tabular}
\end{center}
\caption{Effective potential for temperatures close to the $B=0$ crossover 
temperature.}
\label{fig1}
\end{figure}

In the next section we present explicit results. We~explore the dynamics under a magnetic field 
in a temperature range around the $B=0$~crossover temperature of the model, 
$T_{\rm pc} \simeq 150$~MeV. We~choose this region of temperature because of its 
phenomenological interest in a heavy-ion collision setting. The~LSMq $B=0$ crossover, in the 
mean field approximation, occurs for the parameter values $g=3.3$ and $\lambda = 20$. The 
corresponding (tree-level) vacuum values of the $\sigma$ and quark masses are $m_\sigma 
= 604$~MeV and $m_q = 290$~MeV. With a nonzero~$B$, the chiral transition becomes a first 
order transition, with a critical temperature close to $T^B_c = 180$~MeV; the precise value 
of $T^B_c$ depends on the value of~$B$. 
Since we stay away from such a critical point, these issues do not impact our results.
In connection to the transition temperature, we note that at the mean field level, the 
model does not realize a feature first observed by the lattice simulations of 
Refs.~\cite{Bali:2011qj,Bali:2012zg}, in that the condensate has a nonmonotonic behavior 
as a function~of~$B$ around $T = T_{\rm pc}$. But for temperatures below to $T_{\rm pc}$, 
the LSMq in mean field approximation model does reproduce the qualitative features of the 
lattice results~\cite{Andersen:2014xxa}.

To orientate the discussion of results in the next section, we show in Fig.~\ref{fig1} 
the effective potential $V_{\rm eff}[\sigma_0]$ for $B = 15 m^2_\pi$, Eq.~(\ref{Veff}), 
and $B=0$, and temperatures around~$T = 150$~MeV. The effective potential for zero magnetic 
field, $V^{B=0}_{\rm eff}[\sigma_0]$, is given by~\cite{Nahrgang:2011mg}:
\beq 
V^{B=0}_{\rm eff}[\sigma_0] = U[\sigma_0] - 24 T \int \frac{d^3p}{(2\pi)^3} \, 
\ln\left[1 + e^{-E(\bp)/T}\right], 
\label{Veff-B0}
\eeq 
where $E(\bp) = \sqrt{\bp^2 + m^2_\sigma}$. The figure reveals that 
$|V_{\rm eff}| < |V^{B=0}_{\rm eff}|$ for $|\sigma_0| \le 100$~MeV, a feature 
due to a partial cancellation between $\Omega^{BT}[\sigma_0]$ and $\Omega^{B}[\sigma_0]$, 
with the latter being positive for those values~of~$\sigma_0$.


\section{Dissipation and noise, short-time dynamics}
\label{sec:results}

We start examining the magnetic field impact on the damping coefficient $\eta$, the key quantity 
controlling the fluctuations in the $\sigma$ mean field dynamics. The zero magnetic field $\eta$ 
is given in Ref.~\cite{Rischke:1998qy,Nahrgang:2011mg} for the zero mode only, $\bp =0$, for which 
$E_\sigma \approx \sqrt{\bp^2 + m^2_\sigma} = m_\sigma$:

\beq 
\eta_0 = g^2 \, \frac{2 N_c}{\pi} 
\left[1 - 2n_F(m_\sigma/2) \right] \frac{1}{m^2_\sigma} 
\left(m^2_\sigma - 4m^2_q\right)^{3/2} .
\label{eta0}
\eeq 

\noindent
Putting $E_\sigma = m_\sigma$ and  $\bp_\perp =0$ in Eq.~(\ref{damping-coeff}), one
obtains for the magnetic field dependent damping coefficient:

\beq 
\eta_B = g^2 \, \frac{N_c}{4\pi} \, \left[1 - 2n_F(m_\sigma/2) \right] 
\,\left( eB \right) \, \frac{1}{m^2_\sigma} \sqrt{m^2_\sigma - 4m^2_q}.
\label{etaB}
\eeq

\noindent
We obtain $m_\sigma$ from Eq.~(\ref{sig-mass}). To have a real~$\eta$, we must have 
$m_\sigma > 2 m_q$ in Eqs.~(\ref{etaB}) and (\ref{eta0}), a constraint that reflects 
the kinematical limit for the $\sigma$ decay (at rest) into a quark-antiquark pair, 
$\sigma \rightarrow q\bar{q}$, the only source of dissipation in the model under the
present approximations. We note that $\eta_0 = 0$ for $T < 150$~MeV in this calculation
due to the absence of pions; in the presence of pions, the decay $\sigma \rightarrow 2 
\pi$ leads to a nonzero~$\eta$. 
We recall that our results are valid for strong magnetic 
fields only. Therefore, one cannot extrapolate 
our results to $B=0$; for weak magnetic fields, one needs to use a different 
representation for the magnetized quark propagator, as the LLL approximation is not
valid in this case~\cite{Miransky:2015ava}. But, since weak fields (of strengths 
$\sqrt{eB} \ll \Lambda_{\rm QCD}$) have little impact on chiral properties,
we do not need alternative representations for the quark propagator.

\begin{figure}[h]
\begin{center}
\includegraphics[clip,width=0.50\linewidth]{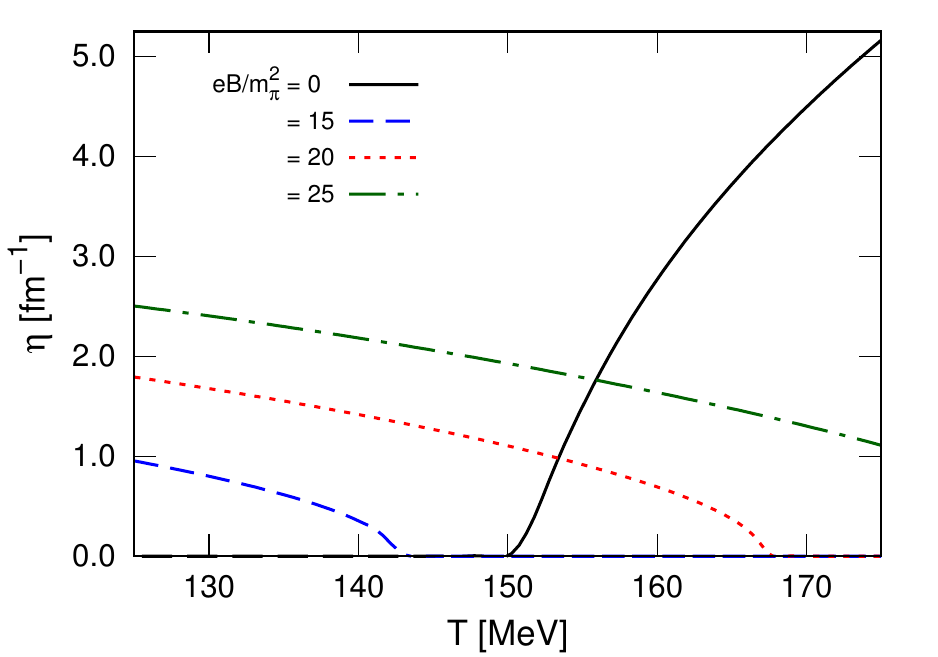}
\end{center}
\caption{Temperature and magnetic field dependence of the zero mode damping coefficient. 
Temperature range chosen to include the $B=0$ pseudocritical temperature, $T_{\rm pc} 
= 150$~MeV. }
\label{fig2}
\end{figure}

Figure~\ref{fig2} displays the temperature dependence of the zero mode damping 
coefficient for $B=0$ and three $B \neq 0$ values. The magnetic field changes the 
qualitative temperature dependence of~$\eta$ close to~$T = 150$~MeV. 
In~a~temperature 
quench scenario, $T \gg T^B_c \rightarrow T \ll T^B_c$, the nonzero value of $\eta_B$ 
for $T < T^B_c$ delays the start off of the condensate evolution after the quench. We 
extend the discussion on this issue at the end of this section, where we study explicit 
short-time solutions of the Langevin equation.

\begin{figure}[h]
\begin{center}
\begin{tabular}{lr}
\includegraphics[clip,width=0.48\linewidth]{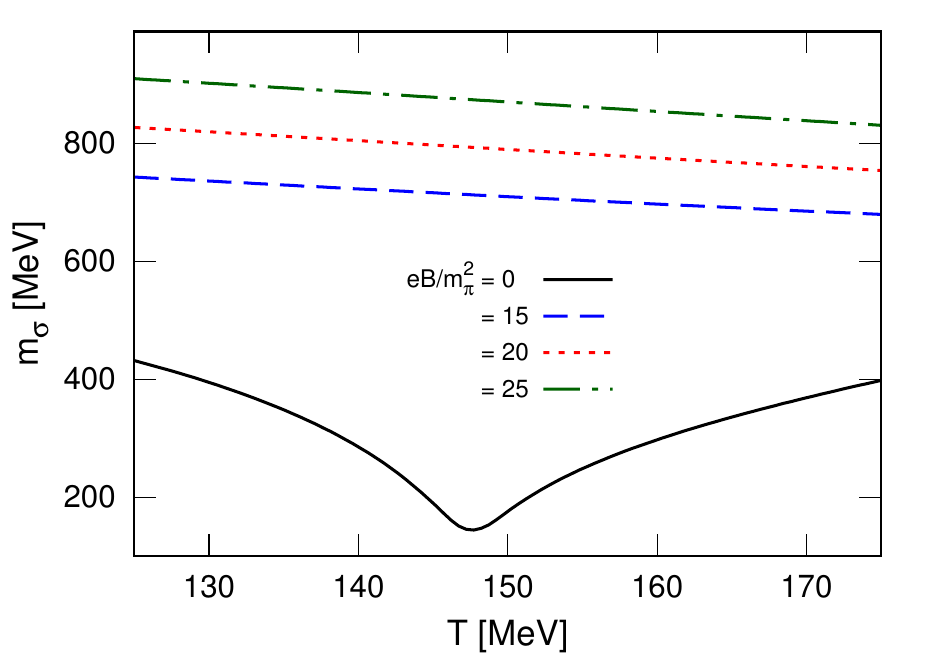} \hspace*{-2.5ex } &
\includegraphics[clip,width=0.48\linewidth]{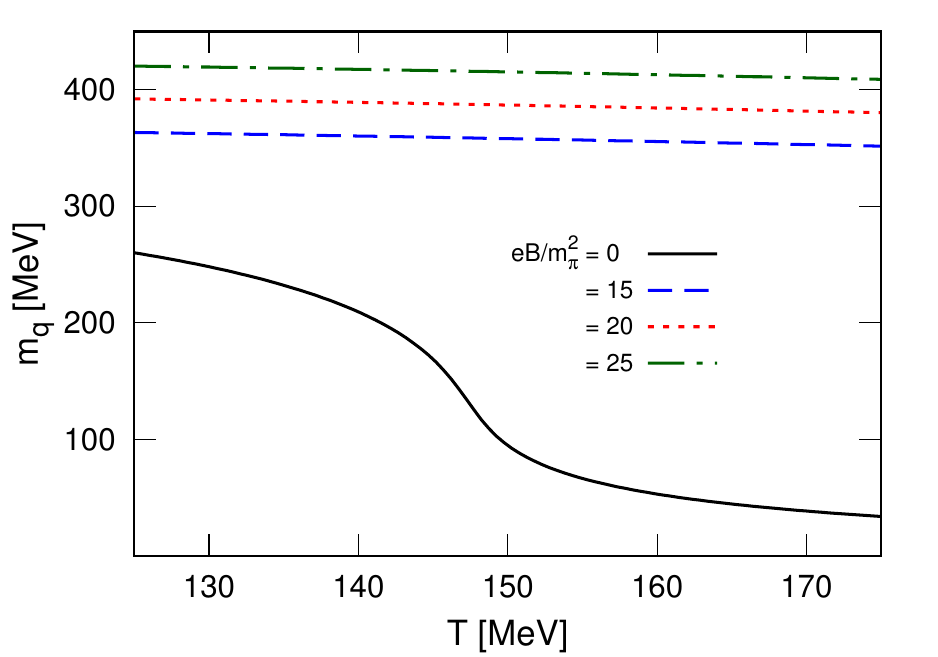}\vspace*{-0ex}
\end{tabular}
\end{center}
\caption{Temperature and magnetic field dependence of the $\sigma$ and quark masses. }
\label{fig3}
\end{figure}

The magnetic field enters the expression for~$\eta$, Eq~(\ref{etaB}), in two ways: through 
the multiplicative $eB$~term, and through the values of $m_\sigma$~and~$m_q$. The latter 
dependence is subtle, $B$~affects $m_\sigma$~and~$m_q$ and thereby affects the inequality 
$m_\sigma > 2 m_q$. 
The magnetic field modifies not only the position of the minimum of $V_{\rm eff}$ 
(which determines $m_q$) but also its curvature around the minimum (which determines
$m_\sigma$){\textemdash}compare the $B\neq 0$ and $B=0$ effective potentials in 
Fig.~1. 
To appreciate this $B$-dependence of $m_\sigma$~and~$m_q$, we show in 
Fig.~\ref{fig3} the temperature dependence of these masses for the values of~$B$ used
in Fig.~\ref{fig2}. It is important 
to notice the different temperature dependence of $m_\sigma$ and $m_q$: the former increases 
faster as the temperature decreases. This faster increase of $m_\sigma$ explains the
$\eta_B$ increase at low temperatures.

Continuing with the aim of gaining analytic understanding, we consider $\sigma$'s dynamics 
in a temperature quench scenario. Before continuing, we spell out the required 
simplifications here. We neglect expansion of the system. Expansion is perhaps 
the most relevant trait of a heavy-ion collision that needs to be taken into account when
simulating a real laboratory event. But such a simulation is out of the scope of this work. 
We also assume a constant magnetic field in the course of the condensate evolution. As such, 
we do not consider the complex magnetohydrodynamics that governs the magnetic field in the 
medium expansion course. The~magnetic field weakens as the system expands, but it also 
induces electric currents that can sustain a magnetic field of sizeable strength while 
the system exists~\cite{Tuchin:2013apa,Gursoy:2014aka,Tuchin:2015oka}. This feature, to 
some extent, justifies the assumption of a constant field. Finally, we do no consider 
reheating, i.e. energy transfer between the condensate and the background. Reheating 
changes the local temperature of the background and, as for zero magnetic fields,
can effect the dynamics~\cite{Nahrgang:2011mv}. We reserve for a separate study the 
inclusion of the neglected effects. 

In a quench scenario, a sudden drop in the temperature 
drives the system out of a high temperature phase, in which $\sigma \approx 0$, and forces 
the system to evolve to a lower temperature phase in which $\sigma \neq 0$. One gets insight 
on how a nonzero $B$ impacts such a quench by examining the time scale controlling the short-time
dynamics. That time scale, which we denote by $\tau_s$, determines how quickly the system 
leaves the initial state. It depends, of course, on $\eta$, and also on the nature of 
the lower temperature phase, which the magnetic field affects as well. This interplay 
between $\eta$ and the nature of the low temperature phase in a quench scenario is well 
known~\cite{Goldenfeld:1992qy,Onuki:2002}. We take as {\em lower temperature phase} one 
around the $B=0$ pseudocritical temperature; that is, at $t=0$ the system is brought 
to one of the local maxima of the $V_{\rm eff}$ in Fig.~\ref{fig1}.  

At short times, when $\sigma \approx 0$, one can linearize the Langevin equation,
neglect the second-order time derivative, and solve the equation analytically. 
It~is~convenient~\cite{Rischke:1998qy} to rescale the fields by 
the volume $V=L^3$, namely $\overline{\sigma} = \sigma/L^3$  and $\overline{\xi}_\sigma 
= \xi_\sigma/L^3$. The~Langevin equation for $\overline{\sigma}$ can be written as:

\beq 
\eta(\bp_\perp) \, \frac{\partial \overline{\sigma}(t,\bp_\perp)}{\partial t} 
- \left(\mu^2 - \bp^2_\perp\right)  \overline{\sigma}(t,\bp_\perp) 
+ g \rho_s(\sigma_0) - f_\pi m^2_\pi 
= \overline{\xi}_{\sigma}(t,\bp_\perp),
\label{EoM-lin0}
\eeq 

\noindent
where 

\beq 
\mu^2 = \lambda \left(f^2_\pi - \frac{m^2_\pi}{\lambda}\right), 
\label{mu}
\eeq 

\noindent
and $\xi_\sigma$ has zero mean, $\langle \xi_\sigma(t,\bp_\perp) \rangle_\xi = 0$,
and correlation

\beq 
\langle \xi_\sigma(t,\bp_\perp) \xi_\sigma(t',\bp'_\perp) \rangle_\xi= 
(2\pi)^2 \delta(\bp_\perp + \bp'_\perp) 
\,L \, 
\delta(t - t') \overline{N}(\bp_\perp) , 
\eeq

\noindent
where ${\overline N}(\bp_\perp) = N(\bp_\perp)/L^6$. We compute the 
equal-time correlation function (variance) of the field, $\langle \, \overline{\sigma}^2(t, \bp^2_\perp) 
\, \rangle_\xi$. Taking as initial condition $\overline{\sigma}(0,\bp_\perp) = 0$, one obtains: 

\bea
\langle \, \overline{\sigma}^2(t, \bp^2_\perp) \, \rangle_\xi &=& 
\frac{\left[g \rho_s(\sigma_0) - f_\pi m^2_\pi\right]^2}
{(\mu^2 - \bp_\perp)^2} \left( e^{ \, \lambda(\bp_\perp) \, t/\tau_s } - 1 \right)^2 
\nn \\[0.35true cm]
&& + \, \frac{E(\bp_\perp) \coth(E(\bp_\perp))}{L^3 (\mu^2 - \bp^2_\perp)} 
\left( e^{\, 2 \, \lambda(\bp_\perp) \, t/\tau_s } - 1 \right),
\label{sig2}
\eea 

\noindent
where 

\beq 
\tau_s = \frac{\eta_B}{\mu^2} \quad \quad \text{and} \quad \quad 
\lambda(\bp_\perp) = \frac{1 - \bp^2_\perp/\mu^2}{\eta(\bp_\perp)/\eta_B},
\label{tau-lambda}
\eeq 

\noindent
with $\eta_B$ given by Eq.~(\ref{etaB}). 

From the definition of $\lambda(\bp_\perp)$ one sees that the exponentials 
in Eq.~(\ref{sig2}) increase with time for long wavelength modes, $\bp^2_\perp < \mu^2$,
and decrease for short wavelengths, $\bp^2_\perp > \mu^2$. That is, long wavelength modes
{\em explode} at short times, akin to the familiar phenomenon of spinodal 
decomposition~\cite{Goldenfeld:1992qy,Onuki:2002}. We~recall that the quench we are 
considering brings the system to one of the local maxima of the effective potential 
$V_{\rm eff}$ in Fig.~\ref{fig1}; there are no barriers to overcome. The {\em explosion} 
is controlled by the time scale $\tau_s$, which depends on $\eta$ (fluctuations) and 
$\mu^2$ (state). The first term in Eq.~(\ref{sig2}) exposes the role played by the low 
temperature phase; it comes from $\delta V_{\rm eff}[\sigma]/\delta\sigma$. Notice that for small
$\sigma$, that term is nothing else $\delta V_{\rm eff}[\sigma]/\delta\sigma = 0$:
$(g \rho_s - f_\pi m^2_\pi)/\mu^2 \equiv \sigma^2_s$, where $\sigma_s$ stands for
{\em small} $\sigma$. The second term comes from the noise source. 

\begin{figure}[h]
\begin{center}
\begin{tabular}{lr}
\includegraphics[clip,width=0.48\linewidth]{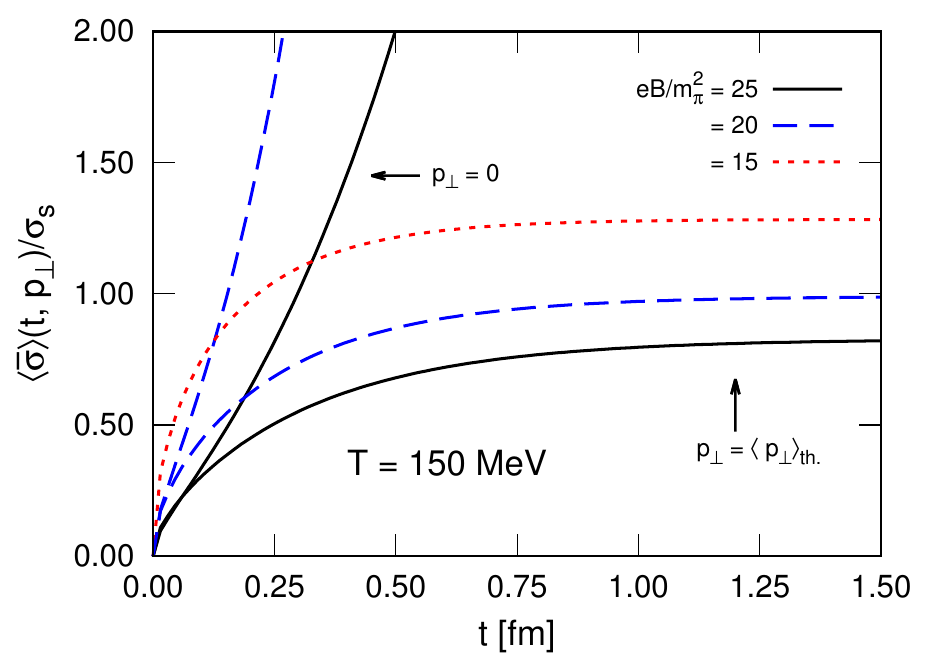} \hspace*{-2.5ex } &
\includegraphics[clip,width=0.48\linewidth]{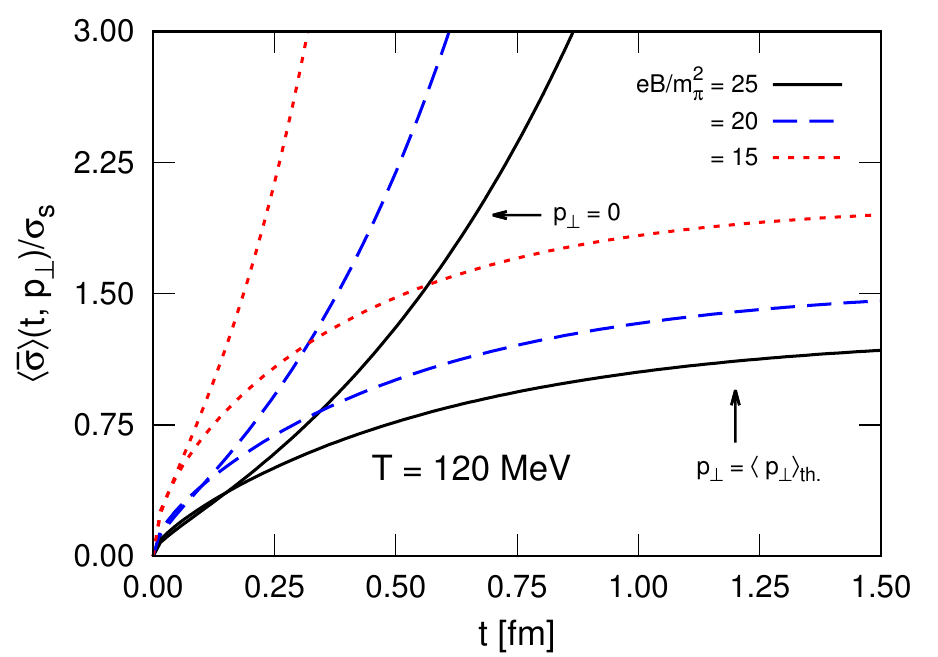}\vspace*{-0ex}
\end{tabular}
\end{center}
\caption{Square root of the equal-time correlation function, normalized to 
$\sigma_s${\textemdash}see text for definitions. Notice the different vertical 
axes ranges in two panels.
There is no dashed-red curve for $\bp_\perp = 0$ in the left panel
because $\eta$ is zero for $eB/m^2_\pi = 15$ and $T = 150$~MeV (see Fig.~\ref{fig2}).}
\label{fig4}
\end{figure}

We present results for the (square root of the) equal-time correlation function for two values 
of $\bp_\perp$; the zero mode $\bp_\perp = 0$, and a thermal average value 
$\langle p_\perp \rangle_{\rm th.} = \sqrt{ \langle \bp^2_\perp \rangle_{\rm th.} }$,
where $\langle \bp^2_\perp \rangle_{\rm th.}$ is the average: 

\beq 
\langle \bp^2_\perp \rangle_{\rm th.} = \frac{\int d^2p_\perp \, p^2_\perp \, n_B(\bp_\perp)}
{\int d^2p_\perp \, n_B(p_\perp)}, \quad \quad \text{where} \quad \quad 
n_B(p) = \frac{1}{ e^{E_\sigma(p)/T} - 1 },
\label{p2-av}
\eeq 

\noindent
with $E_\sigma(p) = \sqrt{p^2 + m^2_\sigma}$. We take
a volume of dimension $L^3 = (10~{\rm fm})^3$.  Figure~\ref{fig4} shows results for 
$\langle \overline{\sigma}\rangle (t,\bp_\perp) / \sigma_s$, 
where we defined $\langle \overline{\sigma} \rangle (t,\bp_\perp) = 
\sqrt{ \langle \, \overline{\sigma}^2(t, \bp_\perp) \, \rangle_\xi}$. The zero mode's 
fast exponential growth stands out in the two panels of the figure. The magnetic field
impact on the short-time growth also stands out, notably the explosion delay alluded
to previously. Since our calculation does not take into account expansion of the
system, it is difficult to assess the phenomenological impact of such a delay,
e.g. on the QGP disassemble into hadrons. However, the delay does not seem
irrelevant in this respect, as it can reach~$1~\rm{fm}$ (right panel of Fig.~\ref{fig4}),
being of the order of $10\%$ of the total time the QGP takes to disassemble 
into hadrons. We recall that the latter is on the average of the order of $10~\rm{fm}$, 
time over which the temperature varies between $T_{\rm ch} \sim 150~{\rm MeV}$ and 
$T_{\rm K} \sim 100~{\rm MeV}$~\cite{Busza:2018rrf,Chatterjee:2015fua}. Here, $T_{\rm ch}$
and $T_{\rm K}$ are respectively the chemical and kinetic freeze out temperatures;
the former signals the end of inelastic collisions and fixes the observed hadron 
abundances and the latter signals the end of elastic hadron collisions and
leads to the disassemble of the system into hadrons. Given that a magnetic field 
also affects hadron masses, there  seems to be room for optimism for a possible 
experimental signal in hadron emission spectra from noncentral collisions. Certainly 
these results warrant further studies. 

Will pions change qualitatively the overall picture? Probably not. For $B=0$, pions   
have a significant effect on the $\sigma$ dynamics only close to the first-order transition
of the LSMq~\cite{weiss2016}. The~results we~have~shown here refer to temperatures away 
from the first order transition temperature, which~we~recall,~$T~\ge~180$~MeV. Moreover, 
results from lattice QCD~\cite{Bali:2017ian} and phenomenological models~\cite{Andersen:2012zc,
Miransky:2015ava,Dumm:2021vop} predict that a background 
magnetic field leaves unchanged the $\pi^0$ mass and increases the $\pi^{\pm}$ masses, 
as expected on general grounds, features that will not change the results of 
Ref.~\cite{weiss2016}. An instance where pions will change quantitatively our results
is in the value of $\eta$: pions bring further dissipation with
the $\sigma \rightarrow 2 \pi$ channel, which implies a positive contribution to
$\tau_s$, i.e. the delay increases. However, the question will be answered only 
with a detailed calculation.


\section{Conclusions and perspectives}
\label{sec:conclusions}

We studied the impact of a strong magnetic field on the chiral quark condensate dynamics.
We~built on the semiclassical framework developed for zero magnetic field developed
in~Ref.~\cite{Nahrgang:2011mg}. That framework bases the dynamics on a mean field 
Langevin equation derived from a microscopic chiral quark model. We extended that
Langevin equation to include the effects of a magnetic field. The~Langevin 
equation we derived features damping and noise modified by the magnetic field. 
Damping and noise reflect the condensate's interactions with an 
effective magnetized quark background in local thermal equilibrium. The background 
results from integrating out quarks from a mean-field effective action defined by 
the linear sigma model. To integrate out quarks, we used the closed time path 
formalism of nonequilibrium quantum field theory. We obtained numerical results using 
values of magnetic field strengths and space-time scales related to high-energy 
heavy-ion collision experiments. We presented results for the short-time condensate 
dynamics under temperature quenches. The~quenches we used were from a high temperature, 
for which the condensate is zero, to lower temperatures close to the zero magnetic field 
crossover temperature, $T \sim 150$~MeV. The results we showed revealed that the magnetic 
field changes the dissipation pattern as compared to the zero magnetic field case, 
retarding condensate's short-time evolution substantially, a feature that can impact 
hadron formation at the QCD transition.

Our study was a first incursion into a complex many-body problem. Our primary aim in
this study was to get insight into how a strong magnetic field affects 
condensate dynamics. We simplified the analysis, and sought an analytical 
understanding whenever possible. We also omitted physical effects peculiar to a 
heavy-ion collision. As such, before one can draw conclusions on phenomenological
consequences in a realist heavy-ion setting, one needs to extend the theoretical
framework to include the omitted features. These include pions, expansion, 
reheating, magnetohydrodynamics modes, and~coupling to other order parameters. 
As in the case of zero magnetic field~\cite{Nahrgang:2011mg}, the formalism 
developed here is flexible enough to tackle the more complex problem.
Another extension of our study is to incorporate a confinement mechanism. 
A possibility is to couple a color dielectric field 
to the chiral $\sigma$ and ${\bm{\pi}}$ fields of the LSMq, 
a possibility very much explored in the context of bag and soliton 
models~\cite{Birse:1991cx}. Such models can be extended to include
explicit gluon degrees of freedom to realize dynamical chiral symmetry breaking
and describe asymptotic freedom~\cite{Krein:1988vh,Krein:1991kz}.

The framework developed in this paper can be adapted to study magnetic field effects
on the QCD phase transition in the early universe and in the interior of magnetized 
compact stars (magnetars). Several mechanisms of strong magnetic field 
generation in the early universe have been suggested~\cite{Vachaspati:1991nm,Grasso:2000wj}; 
a very recent, connected with the QCD phase transition, involves the collapse 
of domain walls related to the confinement order parameter~\cite{Atreya:2017ofd}. 
A~marked difference between the early universe and 
heavy-ion collision settings concerns the rate of change of the temperature $d \ln T/dt$ 
during expansion of the system. In the early universe, this rate is given by Hubble 
constant~$H \sim 10^{-18}~s^{-1}$, which is much slower than that in a heavy-ion collision. Therefore, the primordial chiral condensate evolves in a slowly 
changing effective potential as the system expands. Such an evolution characterizes an 
annealing scenario for the phase change, rather than of a quench, but it can be studied 
equally well with the Langevin equation framework of the present paper~\cite{Gavin:1993px}. 
Regarding magnetars, the inner-core magnetic field can reach strengths varying 
between $eB \simeq m^2_\pi$ and $eB \simeq 50 m^2_\pi$~\cite{Ferrer:2019eof}. 
In this setting, the~temperatures are very low, lower than~50~MeV, and the phase conversion is driven by high 
baryon density. An issue of interest relates to the time scales associated with the
phase conversion during the early stages of the magnetar formation after
a core-collapsing supernova process. In this case, the  
formalism used in this paper needs to be extended to nonzero baryon density~\cite{Kroff:2014qxa}.


\funding{G.K was supported in part by: Conselho Nacional 
de Desenvolvimento Cient\'{\i}fico e Tecnol\'ogico - CNPq, Grants No. 
309262/2019-4, 464898/2014-5 (INCT F\'{\i}sica 
Nuclear e Apli\-ca\-\c{c}\~oes), Funda\c{c}\~ao de Amparo \`a 
Pesquisa do Estado de S\~ao Paulo - FAPESP, Grant No. 2018/25225-9. 
C.M.~was supported by a scholarship from Coordena\c{c}\~ao de Aperfei\c{c}oamento de 
Pessoal de N\'{\i}vel Superior - CAPES.
}

\acknowledgments{G.K. thanks Prof. Ashok Das for e-mail dicussions on the
fermionic CTP Bogoliubov transformation.}

\conflictsofinterest{The authors declare no conflicts of~interest. The funders had no role 
in the design of the study; in the collection, analyses, or interpretation of data; 
in the writing of the manuscript, or in the decision to publish the results.}

\newpage
\abbreviations{The following abbreviations are used in this manuscript.\\
\noindent
\begin{tabular}{ll}
CTP & closed time path \\
LHC & Large Hadron Collider \\
LLL & lowest Landau level \\
LSMq & linear sigma model with quarks  \\
QCD & quantum chromodynamics \\ 
QGP & quark-gluon plasma \\
TFD & thermofield dynamics \\
2PI & two-particle irreducible
\end{tabular}
}

\reftitle{References}


\externalbibliography{yes}
\bibliography{cmd.bib}

\begin{thebibliography}{-------}
\providecommand{\natexlab}[1]{#1}

\bibitem[D'Elia(2013)]{DElia:2012ems}
D'Elia, M.
\newblock {Lattice QCD Simulations in External Background Fields}.
\newblock {\em Lect. Notes Phys.} {\bf 2013}, {\em 871},~181--208,
  \href{http://xxx.lanl.gov/abs/1209.0374}{{\normalfont
  [arXiv:hep-lat/1209.0374]}}.
\newblock
  doi:{\changeurlcolor{black}\href{https://doi.org/10.1007/978-3-642-37305-3_7}{\detokenize{10.1007/978-3-642-37305-3_7}}}.

\bibitem[Endr\"odi(2014)]{Endrodi:2014vza}
Endr\"odi, G.
\newblock {QCD in magnetic fields: from Hofstadter's butterfly to the phase
  diagram}.
\newblock {\em PoS} {\bf 2014}, {\em LATTICE2014},~018,
  \href{http://xxx.lanl.gov/abs/1410.8028}{{\normalfont
  [arXiv:hep-lat/1410.8028]}}.
\newblock
  doi:{\changeurlcolor{black}\href{https://doi.org/10.22323/1.214.0018}{\detokenize{10.22323/1.214.0018}}}.

\bibitem[Ding \em{et~al.}(2020)Ding, Li, Shi, Tomiya, Wang, and
  Zhang]{Ding:2020pao}
Ding, H.T.; Li, S.T.; Shi, Q.; Tomiya, A.; Wang, X.D.; Zhang, Y.
\newblock {QCD phase structure in strong magnetic fields}.
\newblock  {Criticality in QCD and the Hadron Resonance Gas},  2020,
  \href{http://xxx.lanl.gov/abs/2011.04870}{{\normalfont
  [arXiv:hep-lat/2011.04870]}}.

\bibitem[Aoki \em{et~al.}(2006)Aoki, Endrodi, Fodor, Katz, and
  Szabo]{Aoki:2006we}
Aoki, Y.; Endrodi, G.; Fodor, Z.; Katz, S.; Szabo, K.
\newblock {The Order of the quantum chromodynamics transition predicted by the
  standard model of particle physics}.
\newblock {\em Nature} {\bf 2006}, {\em 443},~675--678,
  \href{http://xxx.lanl.gov/abs/hep-lat/0611014}{{\normalfont
  [hep-lat/0611014]}}.
\newblock
  doi:{\changeurlcolor{black}\href{https://doi.org/10.1038/nature05120}{\detokenize{10.1038/nature05120}}}.

\bibitem[Wilczek(2008)]{Wilczek:2008zz}
Wilczek, F.
\newblock {\em {The lightness of being: Mass, ether, and the unification of
  forces}}; Basic Books,  2008.

\bibitem[Roberts(2020)]{Roberts:2020hiw}
Roberts, C.D.
\newblock {Empirical Consequences of Emergent Mass}.
\newblock {\em Symmetry} {\bf 2020}, {\em 12},~1468,
  \href{http://xxx.lanl.gov/abs/2009.04011}{{\normalfont
  [arXiv:hep-ph/2009.04011]}}.
\newblock
  doi:{\changeurlcolor{black}\href{https://doi.org/10.3390/sym12091468}{\detokenize{10.3390/sym12091468}}}.

\bibitem[Vachaspati(1991)]{Vachaspati:1991nm}
Vachaspati, T.
\newblock {Magnetic fields from cosmological phase transitions}.
\newblock {\em Phys. Lett. B} {\bf 1991}, {\em 265},~258--261.
\newblock
  doi:{\changeurlcolor{black}\href{https://doi.org/10.1016/0370-2693(91)90051-Q}{\detokenize{10.1016/0370-2693(91)90051-Q}}}.

\bibitem[Grasso and Rubinstein(2001)]{Grasso:2000wj}
Grasso, D.; Rubinstein, H.R.
\newblock {Magnetic fields in the early universe}.
\newblock {\em Phys. Rept.} {\bf 2001}, {\em 348},~163--266,
  \href{http://xxx.lanl.gov/abs/astro-ph/0009061}{{\normalfont
  [astro-ph/0009061]}}.
\newblock
  doi:{\changeurlcolor{black}\href{https://doi.org/10.1016/S0370-1573(00)00110-1}{\detokenize{10.1016/S0370-1573(00)00110-1}}}.

\bibitem[Kouveliotou \em{et~al.}(1998)Kouveliotou, Dieters, Strohmayer, van
  Paradijs, Fishman, Meegan, Hurley, Kommers, Smith, Frail, and
  Murakami]{Kouveliotou:1998ze}
Kouveliotou, C.; Dieters, S.; Strohmayer, T.; van Paradijs, J.; Fishman, G.J.;
  Meegan, C.A.; Hurley, K.; Kommers, J.; Smith, I.; Frail, D.; Murakami, T.
\newblock {An X-ray pulsar with a superstrong magnetic field in the soft
  gamma-ray repeater SGR 1806-20.}
\newblock {\em Nature} {\bf 1998}, {\em 393},~235--237.
\newblock
  doi:{\changeurlcolor{black}\href{https://doi.org/10.1038/30410}{\detokenize{10.1038/30410}}}.

\bibitem[Duncan and Thompson(1992)]{Duncan:1992hi}
Duncan, R.C.; Thompson, C.
\newblock {Formation of very strongly magnetized neutron stars - implications
  for gamma-ray bursts}.
\newblock {\em Astrophys. J. Lett.} {\bf 1992}, {\em 392},~L9.
\newblock
  doi:{\changeurlcolor{black}\href{https://doi.org/10.1086/186413}{\detokenize{10.1086/186413}}}.

\bibitem[Rafelski and Muller(1976)]{Rafelski:1975rf}
Rafelski, J.; Muller, B.
\newblock {Magnetic Splitting of Quasimolecular Electronic States in Strong
  Fields}.
\newblock {\em Phys. Rev. Lett.} {\bf 1976}, {\em 36},~517.
\newblock
  doi:{\changeurlcolor{black}\href{https://doi.org/10.1103/PhysRevLett.36.517}{\detokenize{10.1103/PhysRevLett.36.517}}}.

\bibitem[Kharzeev \em{et~al.}(2008)Kharzeev, McLerran, and
  Warringa]{Kharzeev:2007jp}
Kharzeev, D.E.; McLerran, L.D.; Warringa, H.J.
\newblock {The Effects of topological charge change in heavy ion collisions:
  'Event by event P and CP violation'}.
\newblock {\em Nucl. Phys. A} {\bf 2008}, {\em 803},~227--253,
  \href{http://xxx.lanl.gov/abs/0711.0950}{{\normalfont
  [arXiv:hep-ph/0711.0950]}}.
\newblock
  doi:{\changeurlcolor{black}\href{https://doi.org/10.1016/j.nuclphysa.2008.02.298}{\detokenize{10.1016/j.nuclphysa.2008.02.298}}}.

\bibitem[Skokov \em{et~al.}(2009)Skokov, Illarionov, and Toneev]{Skokov:2009qp}
Skokov, V.; Illarionov, A.; Toneev, V.
\newblock {Estimate of the magnetic field strength in heavy-ion collisions}.
\newblock {\em Int. J. Mod. Phys. A} {\bf 2009}, {\em 24},~5925--5932,
  \href{http://xxx.lanl.gov/abs/0907.1396}{{\normalfont
  [arXiv:nucl-th/0907.1396]}}.
\newblock
  doi:{\changeurlcolor{black}\href{https://doi.org/10.1142/S0217751X09047570}{\detokenize{10.1142/S0217751X09047570}}}.

\bibitem[Jacak and Muller(2012)]{Jacak:2012dx}
Jacak, B.V.; Muller, B.
\newblock {The exploration of hot nuclear matter}.
\newblock {\em Science} {\bf 2012}, {\em 337},~310--314.
\newblock
  doi:{\changeurlcolor{black}\href{https://doi.org/10.1126/science.1215901}{\detokenize{10.1126/science.1215901}}}.

\bibitem[Shuryak(2017)]{Shuryak:2014zxa}
Shuryak, E.
\newblock {Strongly coupled quark-gluon plasma in heavy ion collisions}.
\newblock {\em Rev. Mod. Phys.} {\bf 2017}, {\em 89},~035001,
  \href{http://xxx.lanl.gov/abs/1412.8393}{{\normalfont
  [arXiv:hep-ph/1412.8393]}}.
\newblock
  doi:{\changeurlcolor{black}\href{https://doi.org/10.1103/RevModPhys.89.035001}{\detokenize{10.1103/RevModPhys.89.035001}}}.

\bibitem[Pasechnik and \v{S}umbera(2017)]{Pasechnik:2016wkt}
Pasechnik, R.; \v{S}umbera, M.
\newblock {Phenomenological Review on Quark\textendash{}Gluon Plasma: Concepts
  vs. Observations}.
\newblock {\em Universe} {\bf 2017}, {\em 3},~7,
  \href{http://xxx.lanl.gov/abs/1611.01533}{{\normalfont
  [arXiv:hep-ph/1611.01533]}}.
\newblock
  doi:{\changeurlcolor{black}\href{https://doi.org/10.3390/universe3010007}{\detokenize{10.3390/universe3010007}}}.

\bibitem[Braun-Munzinger \em{et~al.}(2016)Braun-Munzinger, Koch, Sch\"afer, and
  Stachel]{Braun-Munzinger:2015hba}
Braun-Munzinger, P.; Koch, V.; Sch\"afer, T.; Stachel, J.
\newblock {Properties of hot and dense matter from relativistic heavy ion
  collisions}.
\newblock {\em Phys. Rept.} {\bf 2016}, {\em 621},~76--126,
  \href{http://xxx.lanl.gov/abs/1510.00442}{{\normalfont
  [arXiv:nucl-th/1510.00442]}}.
\newblock
  doi:{\changeurlcolor{black}\href{https://doi.org/10.1016/j.physrep.2015.12.003}{\detokenize{10.1016/j.physrep.2015.12.003}}}.

\bibitem[Goldenfeld(1992)]{Goldenfeld:1992qy}
Goldenfeld, N.
\newblock {\em {Lectures on phase transitions and the renormalization group}};
  Perseus Books: Reading,  1992.

\bibitem[Onuki(2002)]{Onuki:2002}
Onuki, A.
\newblock {\em {Phase Transition Dynamics}}; {Cambridge University Press}:
  Cambridge,  2002.

\bibitem[Rajagopal and Wilczek(1993)]{Rajagopal:1993ah}
Rajagopal, K.; Wilczek, F.
\newblock {Emergence of coherent long wavelength oscillations after a quench:
  Application to QCD}.
\newblock {\em Nucl. Phys. B} {\bf 1993}, {\em 404},~577--589,
  \href{http://xxx.lanl.gov/abs/hep-ph/9303281}{{\normalfont
  [hep-ph/9303281]}}.
\newblock
  doi:{\changeurlcolor{black}\href{https://doi.org/10.1016/0550-3213(93)90591-C}{\detokenize{10.1016/0550-3213(93)90591-C}}}.

\bibitem[Bedaque and Das(1993)]{Bedaque:1993fa}
Bedaque, P.F.; Das, A.K.
\newblock {Out-of-equilibrium phase transitions and a toy model for disoriented
  chiral condensates}.
\newblock {\em Mod. Phys. Lett. A} {\bf 1993}, {\em 8},~3151--3164,
  \href{http://xxx.lanl.gov/abs/hep-ph/9307297}{{\normalfont
  [hep-ph/9307297]}}.
\newblock
  doi:{\changeurlcolor{black}\href{https://doi.org/10.1142/S0217732393002099}{\detokenize{10.1142/S0217732393002099}}}.

\bibitem[Greiner and Muller(1997)]{Greiner:1996dx}
Greiner, C.; Muller, B.
\newblock {Classical fields near thermal equilibrium}.
\newblock {\em Phys. Rev. D} {\bf 1997}, {\em 55},~1026--1046,
  \href{http://xxx.lanl.gov/abs/hep-th/9605048}{{\normalfont
  [hep-th/9605048]}}.
\newblock
  doi:{\changeurlcolor{black}\href{https://doi.org/10.1103/PhysRevD.55.1026}{\detokenize{10.1103/PhysRevD.55.1026}}}.

\bibitem[Biro and Greiner(1997)]{Biro:1997va}
Biro, T.S.; Greiner, C.
\newblock {Dissipation and fluctuation at the chiral phase transition}.
\newblock {\em Phys. Rev. Lett.} {\bf 1997}, {\em 79},~3138--3141,
  \href{http://xxx.lanl.gov/abs/hep-ph/9704250}{{\normalfont
  [hep-ph/9704250]}}.
\newblock
  doi:{\changeurlcolor{black}\href{https://doi.org/10.1103/PhysRevLett.79.3138}{\detokenize{10.1103/PhysRevLett.79.3138}}}.

\bibitem[Rischke(1998)]{Rischke:1998qy}
Rischke, D.H.
\newblock {Forming disoriented chiral condensates through fluctuations}.
\newblock {\em Phys. Rev. C} {\bf 1998}, {\em 58},~2331--2357,
  \href{http://xxx.lanl.gov/abs/nucl-th/9806045}{{\normalfont
  [nucl-th/9806045]}}.
\newblock
  doi:{\changeurlcolor{black}\href{https://doi.org/10.1103/PhysRevC.58.2331}{\detokenize{10.1103/PhysRevC.58.2331}}}.

\bibitem[Xu and Greiner(2000)]{Xu:1999aq}
Xu, Z.; Greiner, C.
\newblock {Stochastic treatment of disoriented chiral condensates within a
  Langevin description}.
\newblock {\em Phys. Rev. D} {\bf 2000}, {\em 62},~036012,
  \href{http://xxx.lanl.gov/abs/hep-ph/9910562}{{\normalfont
  [hep-ph/9910562]}}.
\newblock
  doi:{\changeurlcolor{black}\href{https://doi.org/10.1103/PhysRevD.62.036012}{\detokenize{10.1103/PhysRevD.62.036012}}}.

\bibitem[Fraga and Krein(2005)]{Fraga:2004hp}
Fraga, E.S.; Krein, G.
\newblock {Can dissipation prevent explosive decomposition in high-energy heavy
  ion collisions?}
\newblock {\em Phys. Lett. B} {\bf 2005}, {\em 614},~181--186,
  \href{http://xxx.lanl.gov/abs/hep-ph/0412312}{{\normalfont
  [hep-ph/0412312]}}.
\newblock
  doi:{\changeurlcolor{black}\href{https://doi.org/10.1016/j.physletb.2005.03.079}{\detokenize{10.1016/j.physletb.2005.03.079}}}.

\bibitem[Boyanovsky \em{et~al.}(2006)Boyanovsky, de~Vega, and
  Schwarz]{Boyanovsky:2006bf}
Boyanovsky, D.; de~Vega, H.J.; Schwarz, D.J.
\newblock {Phase transitions in the early and the present universe}.
\newblock {\em Ann. Rev. Nucl. Part. Sci.} {\bf 2006}, {\em 56},~441--500,
  \href{http://xxx.lanl.gov/abs/hep-ph/0602002}{{\normalfont
  [hep-ph/0602002]}}.
\newblock
  doi:{\changeurlcolor{black}\href{https://doi.org/10.1146/annurev.nucl.56.080805.140539}{\detokenize{10.1146/annurev.nucl.56.080805.140539}}}.

\bibitem[Farias \em{et~al.}(2007)Farias, Cassol-Seewald, Krein, and
  Ramos]{Farias:2007xc}
Farias, R.; Cassol-Seewald, N.; Krein, G.; Ramos, R.
\newblock {Nonequilibrium dynamics of quantum fields}.
\newblock {\em Nucl. Phys. A} {\bf 2007}, {\em 782},~33--36,
  \href{http://xxx.lanl.gov/abs/nucl-th/0701074}{{\normalfont
  [nucl-th/0701074]}}.
\newblock
  doi:{\changeurlcolor{black}\href{https://doi.org/10.1016/j.nuclphysa.2006.10.002}{\detokenize{10.1016/j.nuclphysa.2006.10.002}}}.

\bibitem[Nahrgang \em{et~al.}(2011)Nahrgang, Leupold, Herold, and
  Bleicher]{Nahrgang:2011mg}
Nahrgang, M.; Leupold, S.; Herold, C.; Bleicher, M.
\newblock {Nonequilibrium chiral fluid dynamics including dissipation and
  noise}.
\newblock {\em Phys. Rev. C} {\bf 2011}, {\em 84},~024912,
  \href{http://xxx.lanl.gov/abs/1105.0622}{{\normalfont
  [arXiv:nucl-th/1105.0622]}}.
\newblock
  doi:{\changeurlcolor{black}\href{https://doi.org/10.1103/PhysRevC.84.024912}{\detokenize{10.1103/PhysRevC.84.024912}}}.

\bibitem[Nahrgang \em{et~al.}(2012)Nahrgang, Leupold, and
  Bleicher]{Nahrgang:2011mv}
Nahrgang, M.; Leupold, S.; Bleicher, M.
\newblock {Equilibration and relaxation times at the chiral phase transition
  including reheating}.
\newblock {\em Phys. Lett. B} {\bf 2012}, {\em 711},~109--116,
  \href{http://xxx.lanl.gov/abs/1105.1396}{{\normalfont
  [arXiv:nucl-th/1105.1396]}}.
\newblock
  doi:{\changeurlcolor{black}\href{https://doi.org/10.1016/j.physletb.2012.03.059}{\detokenize{10.1016/j.physletb.2012.03.059}}}.

\bibitem[Nahrgang \em{et~al.}(2013)Nahrgang, Herold, Leupold, Mishustin, and
  Bleicher]{Nahrgang:2011vn}
Nahrgang, M.; Herold, C.; Leupold, S.; Mishustin, I.; Bleicher, M.
\newblock {The impact of dissipation and noise on fluctuations in chiral fluid
  dynamics}.
\newblock {\em J. Phys. G} {\bf 2013}, {\em 40},~055108,
  \href{http://xxx.lanl.gov/abs/1105.1962}{{\normalfont
  [arXiv:nucl-th/1105.1962]}}.
\newblock
  doi:{\changeurlcolor{black}\href{https://doi.org/10.1088/0954-3899/40/5/055108}{\detokenize{10.1088/0954-3899/40/5/055108}}}.

\bibitem[Singh \em{et~al.}(2011)Singh, Puri, and Mishra]{Singh2011176}
Singh, A.; Puri, S.; Mishra, H.
\newblock Domain growth in chiral phase transitions.
\newblock {\em Nuclear Physics A} {\bf 2011}, {\em 864},~176 -- 202.
\newblock
  doi:{\changeurlcolor{black}\href{https://doi.org/http://dx.doi.org/10.1016/j.nuclphysa.2011.06.023}{\detokenize{http://dx.doi.org/10.1016/j.nuclphysa.2011.06.023}}}.

\bibitem[Krein(2012)]{Krein:2012zt}
Krein, G.
\newblock {Noise and ultraviolet divergences in the dynamics of the chiral
  condensate in QCD}.
\newblock {\em J. Phys. Conf. Ser.} {\bf 2012}, {\em 378},~012032.
\newblock
  doi:{\changeurlcolor{black}\href{https://doi.org/10.1088/1742-6596/378/1/012032}{\detokenize{10.1088/1742-6596/378/1/012032}}}.

\bibitem[Cassol-Seewald \em{et~al.}(2012)Cassol-Seewald, Farias, Krein, and
  Marques~de Carvalho]{Cassol-Seewald:2012tcq}
Cassol-Seewald, N.; Farias, R.S.; Krein, G.; Marques~de Carvalho, R.
\newblock {Noise and ultraviolet divergences in simulations of
  Ginzburg-Landau-Langevin type of equations}.
\newblock {\em Int. J. Mod. Phys. C} {\bf 2012}, {\em 23},~1240016.
\newblock
  doi:{\changeurlcolor{black}\href{https://doi.org/10.1142/S0129183112400165}{\detokenize{10.1142/S0129183112400165}}}.

\bibitem[Singh \em{et~al.}(2013)Singh, Puri, and Mishra]{Singh201312}
Singh, A.; Puri, S.; Mishra, H.
\newblock Domain growth in chiral phase transitions: Role of inertial dynamics.
\newblock {\em Nuclear Physics A} {\bf 2013}, {\em 908},~12 -- 28.
\newblock
  doi:{\changeurlcolor{black}\href{https://doi.org/http://dx.doi.org/10.1016/j.nuclphysa.2013.03.016}{\detokenize{http://dx.doi.org/10.1016/j.nuclphysa.2013.03.016}}}.

\bibitem[Herold \em{et~al.}(2013)Herold, Nahrgang, Mishustin, and
  Bleicher]{Herold:2013bi}
Herold, C.; Nahrgang, M.; Mishustin, I.; Bleicher, M.
\newblock {Chiral fluid dynamics with explicit propagation of the Polyakov
  loop}.
\newblock {\em Phys. Rev. C} {\bf 2013}, {\em 87},~014907,
  \href{http://xxx.lanl.gov/abs/1301.1214}{{\normalfont
  [arXiv:nucl-th/1301.1214]}}.
\newblock
  doi:{\changeurlcolor{black}\href{https://doi.org/10.1103/PhysRevC.87.014907}{\detokenize{10.1103/PhysRevC.87.014907}}}.

\bibitem[Singh \em{et~al.}(2013)Singh, Puri, and Mishra]{Singh:2013pxa}
Singh, A.; Puri, S.; Mishra, H.
\newblock {Kinetics of phase transitions in quark matter}.
\newblock {\em EPL} {\bf 2013}, {\em 102},~52001,
  \href{http://xxx.lanl.gov/abs/1306.5047}{{\normalfont
  [arXiv:hep-ph/1306.5047]}}.
\newblock
  doi:{\changeurlcolor{black}\href{https://doi.org/10.1209/0295-5075/102/52001}{\detokenize{10.1209/0295-5075/102/52001}}}.

\bibitem[Bluhm \em{et~al.}(2019)Bluhm, Jiang, Nahrgang, Pawlowski, Rennecke,
  and Wink]{Bluhm:2018qkf}
Bluhm, M.; Jiang, Y.; Nahrgang, M.; Pawlowski, J.; Rennecke, F.; Wink, N.
\newblock {Time-evolution of fluctuations as signal of the phase transition
  dynamics in a QCD-assisted transport approach}.
\newblock {\em Nucl. Phys. A} {\bf 2019}, {\em 982},~871--874,
  \href{http://xxx.lanl.gov/abs/1808.01377}{{\normalfont
  [arXiv:hep-ph/1808.01377]}}.
\newblock
  doi:{\changeurlcolor{black}\href{https://doi.org/10.1016/j.nuclphysa.2018.09.058}{\detokenize{10.1016/j.nuclphysa.2018.09.058}}}.

\bibitem[Wu \em{et~al.}(2019)Wu, Wu, and Song]{Wu:2018twy}
Wu, S.; Wu, Z.; Song, H.
\newblock {Universal scaling of the \ensuremath{\sigma} field and net-protons
  from Langevin dynamics of model A}.
\newblock {\em Phys. Rev. C} {\bf 2019}, {\em 99},~064902,
  \href{http://xxx.lanl.gov/abs/1811.09466}{{\normalfont
  [arXiv:nucl-th/1811.09466]}}.
\newblock
  doi:{\changeurlcolor{black}\href{https://doi.org/10.1103/PhysRevC.99.064902}{\detokenize{10.1103/PhysRevC.99.064902}}}.

\bibitem[Calzetta and Hu(2008)]{Calzetta:2008iqa}
Calzetta, E.A.; Hu, B.L.B.
\newblock {\em {Nonequilibrium Quantum Field Theory}}; Cambridge Monographs on
  Mathematical Physics, Cambridge University Press,  2008.
\newblock
  doi:{\changeurlcolor{black}\href{https://doi.org/10.1017/CBO9780511535123}{\detokenize{10.1017/CBO9780511535123}}}.

\bibitem[Bellac(2011)]{Bellac:2011kqa}
Bellac, M.L.
\newblock {\em {Thermal Field Theory}}; Cambridge Monographs on Mathematical
  Physics, Cambridge University Press,  2011.
\newblock
  doi:{\changeurlcolor{black}\href{https://doi.org/10.1017/CBO9780511721700}{\detokenize{10.1017/CBO9780511721700}}}.

\bibitem[Gell-Mann and Levy(1960)]{GellMann:1960np}
Gell-Mann, M.; Levy, M.
\newblock {The axial vector current in beta decay}.
\newblock {\em Nuovo Cim.} {\bf 1960}, {\em 16},~705.
\newblock
  doi:{\changeurlcolor{black}\href{https://doi.org/10.1007/BF02859738}{\detokenize{10.1007/BF02859738}}}.

\bibitem[Fraga and Mizher(2008)]{Fraga:2008qn}
Fraga, E.S.; Mizher, A.J.
\newblock {Chiral transition in a strong magnetic background}.
\newblock {\em Phys. Rev. D} {\bf 2008}, {\em 78},~025016,
  \href{http://xxx.lanl.gov/abs/0804.1452}{{\normalfont
  [arXiv:hep-ph/0804.1452]}}.
\newblock
  doi:{\changeurlcolor{black}\href{https://doi.org/10.1103/PhysRevD.78.025016}{\detokenize{10.1103/PhysRevD.78.025016}}}.

\bibitem[Ayala \em{et~al.}(2009)Ayala, Bashir, Raya, and Sanchez]{Ayala:2009ji}
Ayala, A.; Bashir, A.; Raya, A.; Sanchez, A.
\newblock {Chiral phase transition in relativistic heavy-ion collisions with
  weak magnetic fields: Ring diagrams in the linear sigma model}.
\newblock {\em Phys. Rev. D} {\bf 2009}, {\em 80},~036005,
  \href{http://xxx.lanl.gov/abs/0904.4533}{{\normalfont
  [arXiv:hep-ph/0904.4533]}}.
\newblock
  doi:{\changeurlcolor{black}\href{https://doi.org/10.1103/PhysRevD.80.036005}{\detokenize{10.1103/PhysRevD.80.036005}}}.

\bibitem[Frasca and Ruggieri(2011)]{Frasca:2011zn}
Frasca, M.; Ruggieri, M.
\newblock {Magnetic Susceptibility of the Quark Condensate and Polarization
  from Chiral Models}.
\newblock {\em Phys. Rev. D} {\bf 2011}, {\em 83},~094024,
  \href{http://xxx.lanl.gov/abs/1103.1194}{{\normalfont
  [arXiv:hep-ph/1103.1194]}}.
\newblock
  doi:{\changeurlcolor{black}\href{https://doi.org/10.1103/PhysRevD.83.094024}{\detokenize{10.1103/PhysRevD.83.094024}}}.

\bibitem[Andersen and Khan(2012)]{Andersen:2011ip}
Andersen, J.O.; Khan, R.
\newblock {Chiral transition in a magnetic field and at finite baryon density}.
\newblock {\em Phys. Rev. D} {\bf 2012}, {\em 85},~065026,
  \href{http://xxx.lanl.gov/abs/1105.1290}{{\normalfont
  [arXiv:hep-ph/1105.1290]}}.
\newblock
  doi:{\changeurlcolor{black}\href{https://doi.org/10.1103/PhysRevD.85.065026}{\detokenize{10.1103/PhysRevD.85.065026}}}.

\bibitem[Andersen and Tranberg(2012)]{Andersen:2012bq}
Andersen, J.O.; Tranberg, A.
\newblock {The Chiral transition in a magnetic background: Finite density
  effects and the functional renormalization group}.
\newblock {\em JHEP} {\bf 2012}, {\em 08},~002,
  \href{http://xxx.lanl.gov/abs/1204.3360}{{\normalfont
  [arXiv:hep-ph/1204.3360]}}.
\newblock
  doi:{\changeurlcolor{black}\href{https://doi.org/10.1007/JHEP08(2012)002}{\detokenize{10.1007/JHEP08(2012)002}}}.

\bibitem[Ruggieri \em{et~al.}(2013)Ruggieri, Tachibana, and
  Greco]{Ruggieri:2013cya}
Ruggieri, M.; Tachibana, M.; Greco, V.
\newblock {Renormalized vs Nonrenormalized Chiral Transition in a Magnetic
  Background}.
\newblock {\em JHEP} {\bf 2013}, {\em 07},~165,
  \href{http://xxx.lanl.gov/abs/1305.0137}{{\normalfont
  [arXiv:hep-ph/1305.0137]}}.
\newblock
  doi:{\changeurlcolor{black}\href{https://doi.org/10.1007/JHEP07(2013)165}{\detokenize{10.1007/JHEP07(2013)165}}}.

\bibitem[Fraga \em{et~al.}(2014)Fraga, Mintz, and
  Schaffner-Bielich]{Fraga:2013ova}
Fraga, E.S.; Mintz, B.W.; Schaffner-Bielich, J.
\newblock {A search for inverse magnetic catalysis in thermal quark-meson
  models}.
\newblock {\em Phys. Lett. B} {\bf 2014}, {\em 731},~154--158,
  \href{http://xxx.lanl.gov/abs/1311.3964}{{\normalfont
  [arXiv:hep-ph/1311.3964]}}.
\newblock
  doi:{\changeurlcolor{black}\href{https://doi.org/10.1016/j.physletb.2014.02.028}{\detokenize{10.1016/j.physletb.2014.02.028}}}.

\bibitem[Kamikado and Kanazawa(2014)]{Kamikado:2013pya}
Kamikado, K.; Kanazawa, T.
\newblock {Chiral dynamics in a magnetic field from the functional
  renormalization group}.
\newblock {\em JHEP} {\bf 2014}, {\em 03},~009,
  \href{http://xxx.lanl.gov/abs/1312.3124}{{\normalfont
  [arXiv:hep-ph/1312.3124]}}.
\newblock
  doi:{\changeurlcolor{black}\href{https://doi.org/10.1007/JHEP03(2014)009}{\detokenize{10.1007/JHEP03(2014)009}}}.

\bibitem[Ruggieri \em{et~al.}(2014)Ruggieri, Oliva, Castorina, Gatto, and
  Greco]{Ruggieri:2014bqa}
Ruggieri, M.; Oliva, L.; Castorina, P.; Gatto, R.; Greco, V.
\newblock {Critical Endpoint and Inverse Magnetic Catalysis for Finite
  Temperature and Density Quark Matter in a Magnetic Background}.
\newblock {\em Phys. Lett. B} {\bf 2014}, {\em 734},~255--260,
  \href{http://xxx.lanl.gov/abs/1402.0737}{{\normalfont
  [arXiv:hep-ph/1402.0737]}}.
\newblock
  doi:{\changeurlcolor{black}\href{https://doi.org/10.1016/j.physletb.2014.05.073}{\detokenize{10.1016/j.physletb.2014.05.073}}}.

\bibitem[Ayala \em{et~al.}(2014)Ayala, Hern\'andez, Mizher, Rojas, and
  Villavicencio]{Ayala:2014mla}
Ayala, A.; Hern\'andez, L.A.; Mizher, A.J.; Rojas, J.C.; Villavicencio, C.
\newblock {Chiral transition with magnetic fields}.
\newblock {\em Phys. Rev. D} {\bf 2014}, {\em 89},~116017,
  \href{http://xxx.lanl.gov/abs/1404.6568}{{\normalfont
  [arXiv:hep-ph/1404.6568]}}.
\newblock
  doi:{\changeurlcolor{black}\href{https://doi.org/10.1103/PhysRevD.89.116017}{\detokenize{10.1103/PhysRevD.89.116017}}}.

\bibitem[Ayala \em{et~al.}(2015)Ayala, Loewe, and Zamora]{Ayala:2014gwa}
Ayala, A.; Loewe, M.; Zamora, R.
\newblock {Inverse magnetic catalysis in the linear sigma model with quarks}.
\newblock {\em Phys. Rev. D} {\bf 2015}, {\em 91},~016002,
  \href{http://xxx.lanl.gov/abs/1406.7408}{{\normalfont
  [arXiv:hep-ph/1406.7408]}}.
\newblock
  doi:{\changeurlcolor{black}\href{https://doi.org/10.1103/PhysRevD.91.016002}{\detokenize{10.1103/PhysRevD.91.016002}}}.

\bibitem[Andersen \em{et~al.}(2015)Andersen, Naylor, and
  Tranberg]{Andersen:2014oaa}
Andersen, J.O.; Naylor, W.R.; Tranberg, A.
\newblock {Inverse magnetic catalysis and regularization in the quark-meson
  model}.
\newblock {\em JHEP} {\bf 2015}, {\em 02},~042,
  \href{http://xxx.lanl.gov/abs/1410.5247}{{\normalfont
  [arXiv:hep-ph/1410.5247]}}.
\newblock
  doi:{\changeurlcolor{black}\href{https://doi.org/10.1007/JHEP02(2015)042}{\detokenize{10.1007/JHEP02(2015)042}}}.

\bibitem[Ayala \em{et~al.}(2015)Ayala, Dominguez, Hernandez, Loewe, and
  Zamora]{Ayala:2015lta}
Ayala, A.; Dominguez, C.A.; Hernandez, L.A.; Loewe, M.; Zamora, R.
\newblock {Magnetized effective QCD phase diagram}.
\newblock {\em Phys. Rev. D} {\bf 2015}, {\em 92},~096011,
  \href{http://xxx.lanl.gov/abs/1509.03345}{{\normalfont
  [arXiv:hep-ph/1509.03345]}}.
\newblock [Addendum: Phys.Rev.D 92, 119905 (2015)],
  doi:{\changeurlcolor{black}\href{https://doi.org/10.1103/PhysRevD.92.119905}{\detokenize{10.1103/PhysRevD.92.119905}}}.

\bibitem[Gatto and Ruggieri(2013)]{Gatto:2012sp}
Gatto, R.; Ruggieri, M.
\newblock {Quark Matter in a Strong Magnetic Background}.
\newblock {\em Lect. Notes Phys.} {\bf 2013}, {\em 871},~87--119,
  \href{http://xxx.lanl.gov/abs/1207.3190}{{\normalfont
  [arXiv:hep-ph/1207.3190]}}.
\newblock
  doi:{\changeurlcolor{black}\href{https://doi.org/10.1007/978-3-642-37305-3_4}{\detokenize{10.1007/978-3-642-37305-3_4}}}.

\bibitem[Ayala \em{et~al.}(2015)Ayala, Loewe, Villavicencio, and
  Zamora]{Ayala:2014yla}
Ayala, A.; Loewe, M.; Villavicencio, C.; Zamora, R.
\newblock {On the magnetic catalysis and inverse catalysis of phase transitions
  in the linear sigma model}.
\newblock {\em Nucl. Part. Phys. Proc.} {\bf 2015}, {\em 258-259},~209--212,
  \href{http://xxx.lanl.gov/abs/1409.1517}{{\normalfont
  [arXiv:hep-ph/1409.1517]}}.
\newblock
  doi:{\changeurlcolor{black}\href{https://doi.org/10.1016/j.nuclphysbps.2015.01.045}{\detokenize{10.1016/j.nuclphysbps.2015.01.045}}}.

\bibitem[Miransky and Shovkovy(2015)]{Miransky:2015ava}
Miransky, V.A.; Shovkovy, I.A.
\newblock {Quantum field theory in a magnetic field: From quantum
  chromodynamics to graphene and Dirac semimetals}.
\newblock {\em Phys. Rept.} {\bf 2015}, {\em 576},~1--209,
  \href{http://xxx.lanl.gov/abs/1503.00732}{{\normalfont
  [arXiv:hep-ph/1503.00732]}}.
\newblock
  doi:{\changeurlcolor{black}\href{https://doi.org/10.1016/j.physrep.2015.02.003}{\detokenize{10.1016/j.physrep.2015.02.003}}}.

\bibitem[Andersen \em{et~al.}(2016)Andersen, Naylor, and
  Tranberg]{Andersen:2014xxa}
Andersen, J.O.; Naylor, W.R.; Tranberg, A.
\newblock {Phase diagram of QCD in a magnetic field: A review}.
\newblock {\em Rev. Mod. Phys.} {\bf 2016}, {\em 88},~025001,
  \href{http://xxx.lanl.gov/abs/1411.7176}{{\normalfont
  [arXiv:hep-ph/1411.7176]}}.
\newblock
  doi:{\changeurlcolor{black}\href{https://doi.org/10.1103/RevModPhys.88.025001}{\detokenize{10.1103/RevModPhys.88.025001}}}.

\bibitem[Bjorken and Drell(1965)]{Bjorken:1965zz}
Bjorken, J.D.; Drell, S.D.
\newblock {\em {Relativistic quantum fields}}; McGraw-Hill: New York,  1965.

\bibitem[Koch(1997)]{Koch:1997ei}
Koch, V.
\newblock {Aspects of chiral symmetry}.
\newblock {\em Int. J. Mod. Phys. E} {\bf 1997}, {\em 6},~203--250,
  \href{http://xxx.lanl.gov/abs/nucl-th/9706075}{{\normalfont
  [nucl-th/9706075]}}.
\newblock
  doi:{\changeurlcolor{black}\href{https://doi.org/10.1142/S0218301397000147}{\detokenize{10.1142/S0218301397000147}}}.

\bibitem[Kamenev(2011)]{Kamenev:2011}
Kamenev, A.
\newblock {\em {Field Theory of Non-Equilibrium Systems}}; Cambridge University
  Press,  2011.
\newblock
  doi:{\changeurlcolor{black}\href{https://doi.org/10.1017/CBO9780511721700}{\detokenize{10.1017/CBO9780511721700}}}.

\bibitem[Martin \em{et~al.}(1973)Martin, Siggia, and Rose]{Martin:1973zz}
Martin, P.; Siggia, E.; Rose, H.
\newblock {Statistical Dynamics of Classical Systems}.
\newblock {\em Phys. Rev. A} {\bf 1973}, {\em 8},~423--437.
\newblock
  doi:{\changeurlcolor{black}\href{https://doi.org/10.1103/PhysRevA.8.423}{\detokenize{10.1103/PhysRevA.8.423}}}.

\bibitem[Feynman and Vernon(1963)]{Feynman:1963fq}
Feynman, R.; Vernon, F.L., J.
\newblock {The Theory of a general quantum system interacting with a linear
  dissipative system}.
\newblock {\em Annals Phys.} {\bf 1963}, {\em 24},~118--173.
\newblock
  doi:{\changeurlcolor{black}\href{https://doi.org/10.1016/0003-4916(63)90068-X}{\detokenize{10.1016/0003-4916(63)90068-X}}}.

\bibitem[Loewe and Rojas(1992)]{Loewe:1991mn}
Loewe, M.; Rojas, J.
\newblock {Thermal effects and the effective action of quantum
  electrodynamics}.
\newblock {\em Phys. Rev. D} {\bf 1992}, {\em 46},~2689--2694.
\newblock
  doi:{\changeurlcolor{black}\href{https://doi.org/10.1103/PhysRevD.46.2689}{\detokenize{10.1103/PhysRevD.46.2689}}}.

\bibitem[Elmfors \em{et~al.}(1993)Elmfors, Persson, and
  Skagerstam]{Elmfors:1993wj}
Elmfors, P.; Persson, D.; Skagerstam, B.S.
\newblock {QED effective action at finite temperature and density}.
\newblock {\em Phys. Rev. Lett.} {\bf 1993}, {\em 71},~480--483,
  \href{http://xxx.lanl.gov/abs/hep-th/9305004}{{\normalfont
  [hep-th/9305004]}}.
\newblock
  doi:{\changeurlcolor{black}\href{https://doi.org/10.1103/PhysRevLett.71.480}{\detokenize{10.1103/PhysRevLett.71.480}}}.

\bibitem[Hasan \em{et~al.}(2017)Hasan, Chatterjee, and Patra]{Hasan:2017fmf}
Hasan, M.; Chatterjee, B.; Patra, B.K.
\newblock {Heavy Quark Potential in a static and strong homogeneous magnetic
  field}.
\newblock {\em Eur. Phys. J. C} {\bf 2017}, {\em 77},~767,
  \href{http://xxx.lanl.gov/abs/1703.10508}{{\normalfont
  [arXiv:hep-ph/1703.10508]}}.
\newblock
  doi:{\changeurlcolor{black}\href{https://doi.org/10.1140/epjc/s10052-017-5346-z}{\detokenize{10.1140/epjc/s10052-017-5346-z}}}.

\bibitem[Rath and Patra(2017)]{Rath:2017fdv}
Rath, S.; Patra, B.K.
\newblock {One-loop QCD thermodynamics in a strong homogeneous and static
  magnetic field}.
\newblock {\em JHEP} {\bf 2017}, {\em 12},~098,
  \href{http://xxx.lanl.gov/abs/1707.02890}{{\normalfont
  [arXiv:hep-th/1707.02890]}}.
\newblock
  doi:{\changeurlcolor{black}\href{https://doi.org/10.1007/JHEP12(2017)098}{\detokenize{10.1007/JHEP12(2017)098}}}.

\bibitem[Schwinger(1951)]{Schwinger:1951nm}
Schwinger, J.S.
\newblock {On gauge invariance and vacuum polarization}.
\newblock {\em Phys. Rev.} {\bf 1951}, {\em 82},~664--679.
\newblock
  doi:{\changeurlcolor{black}\href{https://doi.org/10.1103/PhysRev.82.664}{\detokenize{10.1103/PhysRev.82.664}}}.

\bibitem[Das \em{et~al.}(2018)Das, Deshamukhya, Kalauni, and
  Panda]{Das:2018qak}
Das, A.; Deshamukhya, A.; Kalauni, P.; Panda, S.
\newblock {Bogoliubov transformation and the thermal operator representation in
  the real time formalism}.
\newblock {\em Phys. Rev. D} {\bf 2018}, {\em 97},~045015,
  \href{http://xxx.lanl.gov/abs/1801.08097}{{\normalfont
  [arXiv:hep-th/1801.08097]}}.
\newblock
  doi:{\changeurlcolor{black}\href{https://doi.org/10.1103/PhysRevD.97.045015}{\detokenize{10.1103/PhysRevD.97.045015}}}.

\bibitem[Menezes \em{et~al.}(2009)Menezes, Benghi~Pinto, Avancini,
  Perez~Martinez, and Providencia]{Menezes:2008qt}
Menezes, D.P.; Benghi~Pinto, M.; Avancini, S.S.; Perez~Martinez, A.;
  Providencia, C.
\newblock {Quark matter under strong magnetic fields in the Nambu-Jona-Lasinio
  Model}.
\newblock {\em Phys. Rev. C} {\bf 2009}, {\em 79},~035807,
  \href{http://xxx.lanl.gov/abs/0811.3361}{{\normalfont
  [arXiv:nucl-th/0811.3361]}}.
\newblock
  doi:{\changeurlcolor{black}\href{https://doi.org/10.1103/PhysRevC.79.035807}{\detokenize{10.1103/PhysRevC.79.035807}}}.

\bibitem[Farias \em{et~al.}(2014)Farias, Gomes, Krein, and
  Pinto]{Farias:2014eca}
Farias, R.L.S.; Gomes, K.P.; Krein, G.I.; Pinto, M.B.
\newblock {Importance of asymptotic freedom for the pseudocritical temperature
  in magnetized quark matter}.
\newblock {\em Phys. Rev. C} {\bf 2014}, {\em 90},~025203,
  \href{http://xxx.lanl.gov/abs/1404.3931}{{\normalfont
  [arXiv:hep-ph/1404.3931]}}.
\newblock
  doi:{\changeurlcolor{black}\href{https://doi.org/10.1103/PhysRevC.90.025203}{\detokenize{10.1103/PhysRevC.90.025203}}}.

\bibitem[Cassol-Seewald \em{et~al.}(2008)Cassol-Seewald, Copetti, and
  Krein]{CASSOLSEEWALD2008297}
Cassol-Seewald, N.; Copetti, M.; Krein, G.
\newblock Numerical approximation of the Ginzburg–Landau equation with memory
  effects in the dynamics of phase transitions.
\newblock {\em Computer Physics Communications} {\bf 2008}, {\em 179},~297 --
  309.
\newblock
  doi:{\changeurlcolor{black}\href{https://doi.org/https://doi.org/10.1016/j.cpc.2008.03.001}{\detokenize{https://doi.org/10.1016/j.cpc.2008.03.001}}}.

\bibitem[Ebert \em{et~al.}(2000)Ebert, Klimenko, Vdovichenko, and
  Vshivtsev]{Ebert:1999ht}
Ebert, D.; Klimenko, K.G.; Vdovichenko, M.A.; Vshivtsev, A.S.
\newblock {Magnetic oscillations in dense cold quark matter with four fermion
  interactions}.
\newblock {\em Phys. Rev. D} {\bf 2000}, {\em 61},~025005,
  \href{http://xxx.lanl.gov/abs/hep-ph/9905253}{{\normalfont
  [hep-ph/9905253]}}.
\newblock
  doi:{\changeurlcolor{black}\href{https://doi.org/10.1103/PhysRevD.61.025005}{\detokenize{10.1103/PhysRevD.61.025005}}}.

\bibitem[Bali \em{et~al.}(2012{\natexlab{a}})Bali, Bruckmann, Endrodi, Fodor,
  Katz, Krieg, Schafer, and Szabo]{Bali:2011qj}
Bali, G.; Bruckmann, F.; Endrodi, G.; Fodor, Z.; Katz, S.; Krieg, S.; Schafer,
  A.; Szabo, K.
\newblock {The QCD phase diagram for external magnetic fields}.
\newblock {\em JHEP} {\bf 2012}, {\em 02},~044,
  \href{http://xxx.lanl.gov/abs/1111.4956}{{\normalfont
  [arXiv:hep-lat/1111.4956]}}.
\newblock
  doi:{\changeurlcolor{black}\href{https://doi.org/10.1007/JHEP02(2012)044}{\detokenize{10.1007/JHEP02(2012)044}}}.

\bibitem[Bali \em{et~al.}(2012{\natexlab{b}})Bali, Bruckmann, Endrodi, Fodor,
  Katz, and Schafer]{Bali:2012zg}
Bali, G.; Bruckmann, F.; Endrodi, G.; Fodor, Z.; Katz, S.; Schafer, A.
\newblock {QCD quark condensate in external magnetic fields}.
\newblock {\em Phys. Rev. D} {\bf 2012}, {\em 86},~071502,
  \href{http://xxx.lanl.gov/abs/1206.4205}{{\normalfont
  [arXiv:hep-lat/1206.4205]}}.
\newblock
  doi:{\changeurlcolor{black}\href{https://doi.org/10.1103/PhysRevD.86.071502}{\detokenize{10.1103/PhysRevD.86.071502}}}.

\bibitem[Tuchin(2013)]{Tuchin:2013apa}
Tuchin, K.
\newblock {Time and space dependence of the electromagnetic field in
  relativistic heavy-ion collisions}.
\newblock {\em Phys. Rev. C} {\bf 2013}, {\em 88},~024911,
  \href{http://xxx.lanl.gov/abs/1305.5806}{{\normalfont
  [arXiv:hep-ph/1305.5806]}}.
\newblock
  doi:{\changeurlcolor{black}\href{https://doi.org/10.1103/PhysRevC.88.024911}{\detokenize{10.1103/PhysRevC.88.024911}}}.

\bibitem[Gursoy \em{et~al.}(2014)Gursoy, Kharzeev, and
  Rajagopal]{Gursoy:2014aka}
Gursoy, U.; Kharzeev, D.; Rajagopal, K.
\newblock {Magnetohydrodynamics, charged currents and directed flow in heavy
  ion collisions}.
\newblock {\em Phys. Rev. C} {\bf 2014}, {\em 89},~054905,
  \href{http://xxx.lanl.gov/abs/1401.3805}{{\normalfont
  [arXiv:hep-ph/1401.3805]}}.
\newblock
  doi:{\changeurlcolor{black}\href{https://doi.org/10.1103/PhysRevC.89.054905}{\detokenize{10.1103/PhysRevC.89.054905}}}.

\bibitem[Tuchin(2016)]{Tuchin:2015oka}
Tuchin, K.
\newblock {Initial value problem for magnetic fields in heavy ion collisions}.
\newblock {\em Phys. Rev. C} {\bf 2016}, {\em 93},~014905,
  \href{http://xxx.lanl.gov/abs/1508.06925}{{\normalfont
  [arXiv:hep-ph/1508.06925]}}.
\newblock
  doi:{\changeurlcolor{black}\href{https://doi.org/10.1103/PhysRevC.93.014905}{\detokenize{10.1103/PhysRevC.93.014905}}}.

\bibitem[Busza \em{et~al.}(2018)Busza, Rajagopal, and van~der
  Schee]{Busza:2018rrf}
Busza, W.; Rajagopal, K.; van~der Schee, W.
\newblock {Heavy Ion Collisions: The Big Picture, and the Big Questions}.
\newblock {\em Ann. Rev. Nucl. Part. Sci.} {\bf 2018}, {\em 68},~339--376,
  \href{http://xxx.lanl.gov/abs/1802.04801}{{\normalfont
  [arXiv:hep-ph/1802.04801]}}.
\newblock
  doi:{\changeurlcolor{black}\href{https://doi.org/10.1146/annurev-nucl-101917-020852}{\detokenize{10.1146/annurev-nucl-101917-020852}}}.

\bibitem[Chatterjee \em{et~al.}(2015)Chatterjee, Das, Kumar, Mishra, Mohanty,
  Sahoo, and Sharma]{Chatterjee:2015fua}
Chatterjee, S.; Das, S.; Kumar, L.; Mishra, D.; Mohanty, B.; Sahoo, R.; Sharma,
  N.
\newblock {Freeze-Out Parameters in Heavy-Ion Collisions at AGS, SPS, RHIC, and
  LHC Energies}.
\newblock {\em Adv. High Energy Phys.} {\bf 2015}, {\em 2015},~349013.
\newblock
  doi:{\changeurlcolor{black}\href{https://doi.org/10.1155/2015/349013}{\detokenize{10.1155/2015/349013}}}.

\bibitem[Weissenborn-Bresch(2016)]{weiss2016}
Weissenborn-Bresch, S.A.
\newblock {\em On the Impact of Pion Fluctuations on the Dynamics of the Order
  Parameter at the Chiral Phase Transition}; Ruperto-Carola-University of
  Heidelberg: Heidelberg, Germany,  2016.
\newblock
  doi:{\changeurlcolor{black}\href{https://doi.org/https://doi.org/10.11588/heidok.00021599}{\detokenize{https://doi.org/10.11588/heidok.00021599}}}.

\bibitem[Bali \em{et~al.}(2018)Bali, Brandt, Endr\H{o}di, and
  Gl\"a\ss{}le]{Bali:2017ian}
Bali, G.S.; Brandt, B.B.; Endr\H{o}di, G.; Gl\"a\ss{}le, B.
\newblock {Meson masses in electromagnetic fields with Wilson fermions}.
\newblock {\em Phys. Rev. D} {\bf 2018}, {\em 97},~034505,
  \href{http://xxx.lanl.gov/abs/1707.05600}{{\normalfont
  [arXiv:hep-lat/1707.05600]}}.
\newblock
  doi:{\changeurlcolor{black}\href{https://doi.org/10.1103/PhysRevD.97.034505}{\detokenize{10.1103/PhysRevD.97.034505}}}.

\bibitem[Andersen(2012)]{Andersen:2012zc}
Andersen, J.O.
\newblock {Chiral perturbation theory in a magnetic background -
  finite-temperature effects}.
\newblock {\em JHEP} {\bf 2012}, {\em 10},~005,
  \href{http://xxx.lanl.gov/abs/1205.6978}{{\normalfont
  [arXiv:hep-ph/1205.6978]}}.
\newblock
  doi:{\changeurlcolor{black}\href{https://doi.org/10.1007/JHEP10(2012)005}{\detokenize{10.1007/JHEP10(2012)005}}}.

\bibitem[Dumm \em{et~al.}(2021)Dumm, Carlomagno, and Scoccola]{Dumm:2021vop}
Dumm, D.G.; Carlomagno, J.P.; Scoccola, N.N.
\newblock {Strong-interaction matter under extreme conditions from chiral quark
  models with nonlocal separable interactions}.
\newblock {\em Symmetry} {\bf 2021}, {\em 13},~121,
  \href{http://xxx.lanl.gov/abs/2101.09574}{{\normalfont
  [arXiv:hep-ph/2101.09574]}}.
\newblock
  doi:{\changeurlcolor{black}\href{https://doi.org/10.3390/sym13010121}{\detokenize{10.3390/sym13010121}}}.

\bibitem[Birse(1990)]{Birse:1991cx}
Birse, M.C.
\newblock {Soliton models for nuclear physics}.
\newblock {\em Prog. Part. Nucl. Phys.} {\bf 1990}, {\em 25},~1--80.
\newblock
  doi:{\changeurlcolor{black}\href{https://doi.org/10.1016/0146-6410(90)90029-4}{\detokenize{10.1016/0146-6410(90)90029-4}}}.

\bibitem[Krein \em{et~al.}(1988)Krein, Tang, Wilets, and
  Williams]{Krein:1988vh}
Krein, G.; Tang, P.; Wilets, L.; Williams, A.G.
\newblock {Confinement, Chiral Symmetry Breaking and the Pion in a
  Chromodielectric Model of Quantum Chromodynamics}.
\newblock {\em Phys. Lett. B} {\bf 1988}, {\em 212},~362--368.
\newblock
  doi:{\changeurlcolor{black}\href{https://doi.org/10.1016/0370-2693(88)91330-5}{\detokenize{10.1016/0370-2693(88)91330-5}}}.

\bibitem[Krein \em{et~al.}(1991)Krein, Tang, Wilets, and
  Williams]{Krein:1991kz}
Krein, G.; Tang, P.; Wilets, L.; Williams, A.G.
\newblock {The Chromodielectric model: Confinement, chiral symmetry breaking,
  and the pion}.
\newblock {\em Nucl. Phys. A} {\bf 1991}, {\em 523},~548--562.
\newblock
  doi:{\changeurlcolor{black}\href{https://doi.org/10.1016/0375-9474(91)90035-5}{\detokenize{10.1016/0375-9474(91)90035-5}}}.

\bibitem[Atreya and Sanyal(2018)]{Atreya:2017ofd}
Atreya, A.; Sanyal, S.
\newblock {Generation of magnetic fields near QCD Transition by collapsing Z(3)
  domains}.
\newblock {\em Eur. Phys. J. C} {\bf 2018}, {\em 78},~1027,
  \href{http://xxx.lanl.gov/abs/1711.11444}{{\normalfont
  [arXiv:hep-ph/1711.11444]}}.
\newblock
  doi:{\changeurlcolor{black}\href{https://doi.org/10.1140/epjc/s10052-018-6501-x}{\detokenize{10.1140/epjc/s10052-018-6501-x}}}.

\bibitem[Gavin and Muller(1994)]{Gavin:1993px}
Gavin, S.; Muller, B.
\newblock {Larger domains of disoriented chiral condensate through annealing}.
\newblock {\em Phys. Lett. B} {\bf 1994}, {\em 329},~486--492,
  \href{http://xxx.lanl.gov/abs/hep-ph/9312349}{{\normalfont
  [hep-ph/9312349]}}.
\newblock
  doi:{\changeurlcolor{black}\href{https://doi.org/10.1016/0370-2693(94)91094-4}{\detokenize{10.1016/0370-2693(94)91094-4}}}.

\bibitem[Ferrer and Hackebill(2019)]{Ferrer:2019eof}
Ferrer, E.J.; Hackebill, A.
\newblock {Equation of State of a Magnetized Dense Neutron System}.
\newblock {\em Universe} {\bf 2019}, {\em 5},~104.
\newblock
  doi:{\changeurlcolor{black}\href{https://doi.org/10.3390/universe5050104}{\detokenize{10.3390/universe5050104}}}.

\bibitem[Kroff and Fraga(2015)]{Kroff:2014qxa}
Kroff, D.; Fraga, E.S.
\newblock {Nucleating quark droplets in the core of magnetars}.
\newblock {\em Phys. Rev. D} {\bf 2015}, {\em 91},~025017,
  \href{http://xxx.lanl.gov/abs/1409.7026}{{\normalfont
  [arXiv:hep-ph/1409.7026]}}.
\newblock
  doi:{\changeurlcolor{black}\href{https://doi.org/10.1103/PhysRevD.91.025017}{\detokenize{10.1103/PhysRevD.91.025017}}}.

\end{thebibliography}
%



\end{document}